
\input phyzzx
\input epsf
\catcode`\@=11
\normaldisplayskip = 14.4pt plus 3.6pt minus 10.0pt
\normaldispshortskip 6pt plus 5pt
\def\papers{ \papersize\headline=\paperheadline\footline=\paperfootline }
\def\papersize{
\hsize=6.5in \vsize=9in \hoffset=0in \voffset=0in
    \advance\hoffset by \HOFFSET \advance\voffset by \VOFFSET
    \pagebottomfiller=0pc
    \skip\footins=\bigskipamount \normalspace }
\catcode`\@=12
\papers
\lettersize
\overfullrule=0pt
\tolerance=5000
\twelvepoint
\def\npb#1#2#3{{\it Nucl.\ Phys.} {\bf B#1} (19#2) #3}
\def\plb#1#2#3{{\it Phys.\ Lett.} {\bf B#1} (19#2) #3}
\def\mpl#1#2#3{{\it Mod.\ Phys.\ Lett.} {\bf A#1} (19#2) #3}
\def\prl#1#2#3{{\it Phys.\ Rev.\ Lett.} {\bf #1} (19#2) #3}
\def\cmp#1#2#3{{\it Commun.\ Math.\ Phys.} {\bf #1} (19#2) #3}
\def\chapterr#1{\chapter{#1}\message{\the\chapternumber . #1}}
%
\def\insfig#1#2#3#4{\midinsert\vbox to #4truein{\vfil\centerline{
\epsfysize=#4truein\epsfbox{#3}}}
\narrower\narrower\noindent{\bf #1.} #2\endinsert}
%
\def\endli{\hfill\break}
\def\frac#1#2{{#1 \over #2}}
\def\sfrac#1#2{\hbox{$\frac{#1}{#2}$}}
\def\bar#1{\overline{#1}}
\def\p{\partial}

\def\semi{\,\hbox{$\subset\kern-1em\times$}\,}  
\def\vev#1{\langle #1 \rangle}
\def\Vev#1{\left\langle #1 \right\rangle}
\def\bra#1{\langle #1 |}
\def\Bra#1{\left\langle #1 \right|}
\def\ket#1{| #1 \rangle}
\def\Ket#1{\left| #1 \right\rangle}
\def\u#1{{\rm U}(#1)}
\def\su#1{{\rm SU}(#1)}
\def\o#1{{\rm O}(#1)}
\def\so#1{{\rm SO}(#1)}
\def\CA{{\cal A}}
\def\CB{{\cal B}}
\def\CG{{\cal G}}
\def\CH{{\cal H}}
\def\CL{{\cal L}}
\def\CO{{\cal O}}
\def\CP{{\cal P}}
\def\CR{{\cal R}}
\def\CS{{\cal S}}
\def\CT{{\cal T}}
\def\BD{{\bf D}}
\def\BR{{\bf R}}
\def\BZ{{\bf Z}}
\def\ztwo{{{\bf Z}_2}}
\def\wsgroup{{\bf Z}_{2}^{ \,{\rm ws}}}
\def\homotopy{{\bf Z}_2\ast {\bf Z}_2}
\def\classif{B({\bf Z}_2,{\cal G})}
\def\pointph{{\cal G}_{(2)}}
%
\pubnum{hep-th/9404101}
\date{April 1994}
\titlepage
\title{Chern-Simons Gauge Theory on Orbifolds:\break Open Strings from Three
Dimensions}
\author{Petr Ho\v rava\footnote{\ast}{E-mail address:
horava@yukawa.uchicago.edu}\footnote{\star}{Address after September 1, 1994:
Joseph Henry Laboratories, Princeton University, Princeton, New Jersey
08544.}}
\medskip
\address{\centerline{Enrico Fermi Institute}
\centerline{University of Chicago}
\centerline{5640 South Ellis Avenue}
\centerline{Chicago IL 60637, USA}}
\bigskip
\abstract
\baselineskip=16pt
Chern-Simons gauge theory is formulated on three dimensional $\ztwo$
orbifolds.  The locus of singular points on a given orbifold is equivalent to
a link of Wilson lines. This allows one to reduce any correlation function on
orbifolds to a sum of more complicated correlation functions in the simpler
theory on manifolds.  Chern-Simons theory on manifolds is known to be related
to 2D CFT on closed string surfaces; here I show that the theory on orbifolds
is related to 2D CFT of unoriented closed and open string models, i.e.\ to
worldsheet orbifold models.  In particular, the boundary components of the
worldsheet correspond to the components of the singular locus in the 3D
orbifold.  This correspondence leads to a simple identification of the open
string spectra, including their Chan-Paton degeneration, in terms of fusing
Wilson lines in the corresponding Chern-Simons theory.  The correspondence is
studied in detail, and some exactly solvable examples are presented.  Some of
these examples indicate that it is natural to think of the orbifold group
$\ztwo$ as a part of the gauge group of the Chern-Simons theory, thus
generalizing the standard definition of gauge theories.
\endpage
%
\REF\Thurston{W.P. Thurston, ``{\bf The Geometry and Topology of
Three-Manifolds},'' Ch.\ 13 (Princeton University Notes, 1977)}
\REF\DHVW{L. Dixon, J.A. Harvey, C. Vafa and E. Witten, ``Strings on
Orbifolds. I and II,'' \npb{261}{85}{678}; \npb{274}{86}{285}}
\REF\AsOrbs{K.S. Narain, M.H. Sarmadi and C. Vafa, ``Asymmetric Orbifolds,''
\npb{288}{87}{551}}
\REF\wso{P. Ho\v{r}ava, ``Strings on Worldsheet Orbifolds,''
\npb{327}{89}{461}}
\REF\others{A. Sagnotti, in: ``{\bf Non-Perturbative Quantum Field Theory},''
Proceedings of Carg\`{e}se 1987, eds.: G. Mack et al.\ (Plenum Press, 1988)}
\REF\DVVV{R. Dijkgraaf, C. Vafa, E. Verlinde and H. Verlinde, ``The Operator
Algebra of Orbifold Models,'' \cmp{123}{89}{485}}
\REF\WSOdual{P. Ho\v{r}ava, ``Background Duality of Open String Models,''
\plb{231}{89}{251}}
\REF\bh{P. Ho\v rava, ``Two Dimensional Stringy Black Holes with One
Asymptotically Flat Domain,'' \plb{289}{92}{293}}
\REF\WJones{E. Witten, ``Quantum Field Theory and the Jones Polynomial,''
\cmp{121}{89}{351}}
\REF\Wsearch{E. Witten, ``The Search for Higher Symmetry in String Theory,''
{\it Phil.\ Trans.\ Roy.\ Soc.}\ {\bf A 329} (1989) 345}
\REF\etsm{P. Ho\v rava, ``Equivariant Topological Sigma Models,''
Prague Inst.\ of Phys.\ preprint PRA-HEP-90/18 (December 1990),
hep-th/9309124, to appear in {\it Nucl.\ Phys.} {\bf B} (1994)}
\REF\cswold{P. Ho\v rava, ``Open Strings from Three Dimensions:
Chern-Simons-Witten Theory on Orbifolds,'' Prague Inst.\ of Phys.\ preprint
PRA-HEP-90/3 (July 1990)\endli
cf.\ also: P. Ho\v rava, ``Orbifold Approach to Open String Theory,'' Prague
Inst.\ of Phys.\ PhD Thesis (1991), pp.\ 57-104}
\REF\affleck{I. Affleck, ``Conformal Field Theory Approach to Quantum Impurity
Problems,'' Vancouver preprint UBCTP-93-25 (November 1993)}
\REF\cklm{C.G. Callan, I.R. Klebanov, A.W.W. Ludwig and J.M. Maldacena,
``Exact Solution of a Boundary Conformal Field Theory,'' Princeton/IAS
preprint PUPT-1450=IASSNS-HEP-94/15 (February 1994)}
\REF\Wstat{E. Witten, ``Gauge Theories and Integrable Lattice Models,''
\npb{322}{89}{351}}
\REF\Oxford{``{\bf Oxford Seminars on Jones-Witten Theory},'' Oxford Notes,
December 1988}
\REF\atiyahbook{M. Atiyah, ``{\bf The Geometry and Physics of Knots}''
(Cambridge U. Press, 1990)}
\REF\DijkW{R. Dijkgraaf and E. Witten, ``Topological Gauge Theories and Group
Cohomology,'' \cmp{129}{90}{393}}
\REF\MSZoo{G. Moore and N. Seiberg, ``Taming the Conformal Zoo,''
\plb{220}{89}{422}}
\REF\EMSS{S. Elitzur, G. Moore, A. Schwimmer and N. Seiberg, ``Remarks on the
Canonical Quantization of the Chern-Simons-Witten Theory,'' \npb{326}{89}{108}}
\REF\CFTag{L. Alvarez-Gaum\'{e}, G. Sierra and C. Gomez, ``Topics in Conformal
Field Theory,'' in: ``{\bf Physics and Mathematics of Strings},'' the Vadim
Knizhnik Memorial Volume, eds.\ L. Brink, D. Friedan and A.M. Polyakov (World
Scientific, 1990)}
\REF\bc{C.G. Callan, C. Lovelace, C.R. Nappi and S.A. Yost, ``Adding Holes and
Crosscaps to the Superstring,'' \npb{293}{87}{83}
\endli J. Polchinski and Y. Cai, ``Consistency of Open Superstring Theories,''
\npb{296}{88}{91}}
\REF\Wgrav{E. Witten, ``2+1 Dimensional Gravity as an Exactly Soluble
System,'' \npb{311}{88}{46}}
\REF\Kogan{Ya.I. Kogan, ``The Off-Shell Closed Strings as Topological Open
Membranes.  Dynamical Transmutation of World-Sheet Dimension,''
\plb{231}{89}{377}}
\REF\CarlipKogan{S. Carlip and Ya.\ Kogan, ``Quantum Geometrodynamics of the
Open Topological Membrane and String Moduli Space,'' \prl{64}{90}{1487}}
\REF\HVer{H. Verlinde, ``Conformal Field Theory, 2-D Quantum Gravity and
Quantization of Teichm\"uller Space,'' \npb{337}{90}{652}}
\REF\Aframings{M.F. Atiyah, ``On Framings of 3-Manifolds,'' {\it Topology}
{\bf 29} (1990) 1}
\REF\Carlip{S. Carlip, ``Exact Quantum Scattering in 2+1 Dimensional
Gravity,'' \npb{324}{89}{106}}
\REF\Cardy{J.L. Cardy, ``Boundary Conditions, Fusion Rules, and the Verlinde
Formula,'' \npb{324}{89}{581}}
\REF\Ishib{N. Ishibashi, ``The Boundary and Crosscap States in Conformal Field
Theories,'' \mpl{4}{89}{251}}
\REF\Dieck{T. tom Dieck, ``Faserb\"undel mit Gruppenoperation,'' {\it Arch.\
Math.} {\bf 20} (1969) 136\endli
see also: T. tom Dieck, ``{\bf Transformation Groups},'' de~Gruyter Series in
Math.\ {\bf 8} (W.~de~Gruyter, Berlin 1987)}
\REF\Wtrans{E. Witten, ``Topology-Changing Amplitudes in 2+1 Dimensional
Gravity,'' \npb{323}{89}{113}}
\REF\KWilcz{L.M. Krauss and F. Wilczek, ``Discrete Gauge Symmetry in Continuum
Theories,'' \prl{63}{89}{1221}}
\REF\Spanier{E.H. Spanier, ``{\bf Algebraic Topology}'' (McGraw Hill,
New York, 1966)}
\REF\CFloyd{P.E. Conner and E.E. Floyd, ``{\bf Differentiable Periodic Maps}''
(Springer Verlag, Berlin, 1964)}
\REF\Stong{R.E. Stong, ``Unoriented Bordism and Actions of Finite Groups,''
{\it Mem.\ AMS} {\bf 103} (1970) 1}
\REF\Bredon{G.E. Bredon, ``{\bf Equivariant Cohomology Theories},'' {\it
Lecture Notes in Math.}\ {\bf 34} (Springer Verlag, Berlin, 1967)}
\REF\BottTu{R. Bott and L.W. Tu, ``{\bf Differential Forms in Algebraic
Geometry}'' (Springer Verlag, Berlin, 1982)}
%
%
\chapterr{Introduction}

Since the first appearance of the  notion of ``orbifolds'' in Thurston's
1977 lectures on three dimensional topology [\Thurston], orbifolds have become
very appealing objects for physicists.  This interest was mainly motivated
by the fact that orbifold singularities are so mild that strings can propagate
consistently on orbifold targets, without violating unitarity of the string
S~matrix [\DHVW].

In critical string theory, orbifolds as targets for string propagation have
been generalized to more subtle structures.  One example are asymmetric
orbifolds [\AsOrbs], i.e.\ two-dimensional (2D) conformal field theories
(CFTs) in which left-movers and right-movers take values in distinct
orbifolds.  The geometrical structure of these generalizations becomes more
involved:  Intriguing subtleties come into play, related in particular to the
geometry of fixed points and the vacuum degeneracy of twisted sectors
[\AsOrbs].  Another generalization is given by worldsheet orbifolds [\wso]
(see also [\others,\DVVV] for related points of view), where string theoretic
vacua are orbifolded by a symmetry that acts directly on the 2D CFTs
describing the vacua.  It is within this construction that open strings emerge
as twisted states of an orbifold.  Since the orbifold interpretation of open
string models first arose, the technique has been useful in several instances
where one is interested in the open string counterparts of closed string
constructions (such as target duality [\WSOdual] or 2D black holes [\bh]).

While now we understand fairly well that open strings come from twisted
sectors in a class of generalized orbifold models, the geometry of this
generalization is not yet completely understood.  Indeed, the target orbifold
geometry gets here even more entangled with the structure of the conformal
field theory itself.  On the other hand, one not fully understood issue in
open string theory is the degeneration of the ground state in the open string
sector of a given model.  Traditionally, this degeneration is constructed by
the Chan-Paton mechanism, which eventually leads to the presence of
non-abelian Yang-Mills gauge symmetry in the spacetime theory.  In the
Chan-Paton mechanism, the degeneration is caused by the somewhat {\it ad hoc}
procedure of inserting charges of a spacetime gauge group (typically $\so N$)
at the ends of open strings.  The seeming arbitrariness in the choice of the
gauge group is eventually fixed by the check of BRST invariance of the string
model, which leads to an essentially unique gauge symmetry group for each
model.  This state of affairs seems unsatisfactory, and a deeper explanation
of the existence and of the high degree of uniqueness of the Chan-Paton
mechanism is to be sought.  In fact, some hints are offered by the orbifold
construction of open string theory:  Open strings belong to twisted sectors on
orbifolds, and one may expect connections between their vacuum degeneracy and
some sort of generalized fixed-point geometry of the orbifold
[\wso,\WSOdual].  I will clarify some of these subtleties in this paper,
making use of a higher-dimensional perspective.

Let us now leave the stringy intricacies aside and consider something simpler,
namely a quantum field theory on orbifolds.  Doing this, however, one
typically encounters inconsistencies:  the scattering S~matrix of local
excitations is not unitary, as particles may leave the world through the
singular points.  There is, however, one important loophole in this argument.
Were we considering a topological quantum field theory, there would be no
local excitations, no S~matrix, and hence no violation of the S~matrix
unitarity.  Thus, we are free to construct a quantum field theory on
orbifolds, on condition that the theory has no local excitations, i.e.\ that
it is topological.

In this paper, we will be concerned with Chern-Simons (CS) gauge theory
[\WJones] on three dimensional orbifolds.  One motivation for this is string
theoretical.  Below I will argue that CS gauge theory on three dimensional
$\ztwo$ orbifolds is related to the theory of open strings in precisely the
same sense as CS gauge theory on manifolds is related to the theory of closed
strings (or more precisely, to rational CFT on compact oriented surfaces of
closed string theory).  This will give us the higher-dimensional perspective
of the puzzles of open string theory that I mentioned above.  The 3D vantage
point as a tool explaining various properties of 2D CFT has been advocated by
Witten [\Wsearch] in the context of CFTs on closed oriented Riemann surfaces;
it is the open-string extension of this ideology that is new in this paper.
Another motivation for the present work may come from the fact that the
Chern-Simons gauge theory on orbifolds represents an explicit example of
equivariant topological quantum field theory in the sense of the axiomatics
presented in [\etsm].

This paper is organized as follows.  In \S{2} I fix notation and review some
aspects of 2D CFT of worldsheet orbifolds and their relation to open strings.
In particular, the structure of possible group actions that generate open
strings in these orbifolds models is elucidated.  I also review briefly some
basic aspects of the Chern-Simons gauge theory on manifolds, in particular its
connection with CFTs on closed oriented Riemann surfaces.  In \S{2.2} it is
shown how, upon looking for a 3D description of worldsheet orbifold CFTs, we
are led to CS gauge theory on $\ztwo$ orbifolds.  This allows us to make some
preliminary conjectures about the correspondence between the spectra of these
two theories.

These conjectures are confirmed in the remainder of the paper, where
quantization of CS gauge theory on orbifolds is analyzed and  a set of
specific examples is given.  In \S{3} I discuss the quantum Chern-Simons gauge
theory on orbifolds, first for arbitrary connected, simply connected gauge
group $G$, and specializing to $G=\su 2$ afterwards.  For any $\ztwo$
orbifold, the locus of all singular points comprises a link in the underlying
topological manifold.  Inside correlation functions, the singular locus is
equivalent to a link of Wilson lines, which allows us to reduce the theory on
orbifolds to a related theory on manifolds.  This theory on manifolds is not
necessarily the CS gauge theory with the same gauge group, as will be seen in
detail in \S{3}.  The question of framing of the components of the singular
locus, raised by their interpretation as a sum of Wilson lines, is studied
briefly in \S{3.2}.  I complete the basic setting for the quantum theory on
orbifolds in \S{3.3}, where I discuss skein theory for the singular locus, and
in \S{3.4}, where the issue of observables is analyzed.

In the beginning of \S{4} I discuss the correspondence between CS gauge theory
on $\ztwo$ orbifolds on the one hand, and 2D CFT of worldsheet orbifolds on
the other.  Most remarkably, the structure of Chan-Paton factors is elucidated
(and fixed uniquely) within CS gauge theory in terms of the algebraic geometry
of the singular locus. \S{}\S{4} and 5 offer a set of basic examples that
illustrate the correspondence.   In \S{4} I study the CFT/CS gauge theory
relation for $\su 2$, while in \S{5} the set of examples is extended to $c=1$
CFTs (corresponding to $G=\u 1$ CS gauge theory), and to holomorphic orbifold
CFTs (CS gauge theory with discrete gauge groups).  Those worldsheet
orbifolds whose orbifold group mixes nontrivially the worldsheet parity
transformation with a target action (the so-called ``exotic worldsheet
orbifolds'') are shown to lead to an unusual form of gauge theory in 3D in
which the orbifold group $\ztwo$ is mixed nontrivially with the CS gauge
group.  Possible implications of this phenomenon are discussed briefly in
\S{5.3}.  Some elements of orbifold topology and geometry that are needed for
the body of the paper are gathered in Appendix A; some more involved
mathematical aspects of the definition of the Lagrangian for CS gauge theory
on orbifolds with general gauge groups are deferred to Appendix B.

This paper is a rewritten version of a paper that was published in July 1990
as a Prague Institute of Physics preprint [\cswold].  Although the results
presented here are the same as in [\cswold], the presentation has been
altered.  A part of the motivation for this revision (apart from the interest
in the theory for reasons discussed above) comes from the possible
applications this theory may have to boundary scattering in 1+1 dimensional
CFT.  In fact, this recently very active area has a remarkably broad domain of
applications, ranging from quantum impurity problems (such as the Kondo
effect) to dissipative quantum mechanics, to propagation in quantum wires, to
the Callan-Rubakov effect, to quantum theory of black holes.  (See [\affleck]
and references therein for a review of most of these applications.)  In many
of these cases, the S~matrix of the boundary scattering exhibits interesting
properties [\affleck,\cklm] whose explanation, I believe, could come from
the correspondence between 2D CFT on surfaces with boundaries and 3D
Chern-Simons theory on $\ztwo$ orbifolds as discussed in this paper.  In fact,
this correspondence suggests that the boundary scattering in 2D CFT can be
alternatively described as an Aharonov-Bohm effect in 3D Chern-Simons gauge
theory.  I hope to return to this point elsewhere.

%
\chapterr{Chern-Simons Gauge Theory on 3D Orbifolds}

Chern-Simons gauge theory was formulated by Witten in [\WJones] as a gauge
theory in three dimensions with compact, connected and simply connected gauge
group $\CG$, and with the Lagrangian given by the Chern-Simons functional,
\foot{Our normalization of $S$ is such that the functional integral is
weighted by $e^{iS}$.}
$$S(A)=\frac{k}{4\pi}\int _M {\rm Tr} \left( A\wedge dA+\sfrac{2}{3}A\wedge
A\wedge A \right).\eqn\action
$$

The set of observables of the theory is generated by Wilson lines
$$
W_R(C)={\rm Tr}_R\,P\exp \int _CA,\eqn\eeobs
$$
where $R$ belong to the finite set of integrable representations of the
Kac-Moody algebra $\widehat\CG$ at level $k$, and $C$ is a closed line in
$M$; and by ``baryon'' configurations first introduced in [\Wstat] and defined
using trivalent vertices.  At the quantum level, only a finite number of these
vertices are relevant, corresponding to the information encoded in the
structure constants of the fusion algebra of the associated WZW model.

Thus, the natural things to calculate are the correlation functions of the
objects just mentioned:
$$\Vev{W_{R_1}(C_1)\,W_{R_2}(C_2)\cdots}_M\equiv \int DA\;W_{R_1}(C_1)\,
W_{R_2}(C_2)\cdots e^{iS(A)}.\eqn\eecorr
$$

A particularly natural way of computing these correlation functions is the
canonical quantization approach.  To any surface $\Sigma$ pierced in points
$z_i$ by Wilson lines in representations $R_i$ there corresponds a
(finite-dimensional) Hilbert space of quantum states, $\CH_{\Sigma,R_i}$.
Cutting the 3D manifold $M$ into two parts along an orientable surface
$\Sigma$, we can compute the amplitude as an inner product within
$\CH_\Sigma$, making use of the fact that the theory satisfies the axioms of
topological QFT [\Oxford--\DijkW].

The key to the appeal of CS gauge theory for string physicists lies in the
elegant relation of the theory to 2D rational CFT [\WJones,\MSZoo,\EMSS]
(for a review of 2D CFT, see e.g.\ [\CFTag]).  This correspondence identifies
the Hilbert space of CS gauge theory canonically quantized on
$\Sigma\times\BR$, where $\Sigma$ is a closed oriented surface, with the
space of all conformal blocks of a rational CFT on $\Sigma$
[\WJones,\MSZoo,\EMSS].  Since in this paper we are mainly interested in the
correspondence between CS gauge theory and CFT, and this correspondence is
well understood only for rational CFTs/compact Chern-Simons gauge groups, we
will restrict ourselves to rational CFTs throughout the paper, without
mentioning the word ``rational'' explicitly.

\section{Conformal Field Theory of Worldsheet Orbifolds}

Our central interest throughout this paper is to reproduce two dimensional CFT
of worldsheet orbifolds from CS gauge theory.  I believe that a short review
of the theory of worldsheet orbifolds may be useful.  For other results not
gathered here, see [\wso,\WSOdual].

Let us choose a left-right symmetric CFT.  Assume also that there is a
discrete group $\widetilde G$ acting as a symmetry group on the theory in the
target, i.e.\  exactly as in [\DVVV].  The theory is by assumption
parity-invariant, i.e.\ there is a symmetry action of the worldsheet
transformation
$$\Omega _0 :(z,\bar z)\mapsto (e^{2\pi i}\bar z,e^{-2\pi i}z)\eqn\eeome$$
on the fields of the theory.  This particular action of the $\ztwo$ group on
the 2D theory (i.e.\ the action that reverses the orientation of the
worldsheet) plays a central role in the paper, and deserves a special
notation;  from now on, I will denote by $\wsgroup$ this particular $\ztwo$
group generated by $\Omega_0$ (or more precisely, by $\Omega$, which is
$\Omega_0$ lifted trivially to the fields of the 2D theory).

Worldsheet orbifolds are then defined as orbifolds whose orbifold group $G$
combines the worldsheet action of $\wsgroup$ with a target symmetry given by
$\widetilde G$, i.e.\
$$G\subset\widetilde G\times\wsgroup.\eqn\eecomb$$
On worldsheet orbifolds, we can get essentially two distinct classes of
twists.  First, if $G$ contains elements of the form $\widetilde g\times 1$,
where 1 is the identity of $\wsgroup$ and $\widetilde g$ is in $\widetilde G$,
then usual twisted states are produced, exactly as in traditional (target)
orbifold models.  The other possibility, i.e.\ the case of twisting by an
element acting non-trivially on the worldsheet by $\Omega$, is a bit more
intricate.  In this case, we can easily observe that the choice of just one
twisting element of $G$, say $g_1\times \Omega$ (where $g_1$ is in $\widetilde
G$), is not sufficient to fully determine the twisted state.  If $g_1\times
\Omega$ corresponds to the twist of fields when we go around the cylindrical
worldsheet in one direction,
$$\phi(e^{2\pi i}z,e^{-2\pi i}\bar z)=(g_1\times\Omega)\cdot\phi(z,\bar z)
\equiv g_1\cdot\phi (e^{2\pi i}\bar z,e^{-2\pi i}z),\eqn\twistplus
$$
we have to add another element, say $g_2\times \Omega$, to determine the twist
in the opposite direction:
$$\phi(e^{-2\pi i}z,e^{2\pi i}\bar z)=(g_2\times\Omega)\cdot\phi(z,\bar z)
\equiv g_2\cdot\phi(e^{2\pi i}\bar z,e^{-2\pi i}z).\eqn\twistminus
$$
It is easy to show that \twistplus\ and \twistminus\ lead to open string
sectors [\wso].

This unusual structure of twisted states has a natural explanation if we think
of the state twisted by the couple $g_1\times\Omega$, $g_2\times\Omega$ as an
open string state, with the open string being a $\ztwo$ orbifold of the closed
string.  To specify twists on a particular worldsheet $\Sigma$, we have to
specify monodromies of fields on $\Sigma$, i.e.\ a representation of the first
homotopy group of $\Sigma$ in the orbifold group:
$$\pi _1(\Sigma )\rightarrow G.\eqn\eehomot$$
The open string is topologically an orbifold $\CO_S$ of the closed string
$S^1$ by $\ztwo$, $\CO_S\equiv S^1/\ztwo$, and its orbifold fundamental group
(see [\Thurston] for the definition) is $\BD$, the infinite dihedral group:
$$\pi _1(\CO_S )=\BD\equiv\ztwo\ast\ztwo\equiv\ztwo\semi\BZ.\eqn\pione$$
Here $\ast$ denotes the free product of groups, and $\semi$ is the
semi-direct product.  The monodromy of the open sector corresponds to a
representation of the first homotopy group of the open string in the orbifold
group:
$$\homotopy \rightarrow G,\eqn\reptwist$$
required to satisfy one obvious geometrical constraint.  The fundamental group
of the open string, $\homotopy$, is naturally mapped onto the group
$\wsgroup$, both of its $\ztwo$ factors being mapped isomorphically to
$\wsgroup$.  Moreover, the orbifold group $G$ has, as a natural subgroup in
$\widetilde G\times\wsgroup$, a canonical projection onto $\wsgroup$.  The
worldsheet orbifold with the orbifold group $G$ then admits only those
representations \reptwist\ that complete the diagram
$$\homotopy \rightarrow \wsgroup \leftarrow G\eqn\tobetriangle$$
to a commutative triangle.  If $G$ itself has the structure of a product:
$$G=G_{0}\times \wsgroup ,\eqn\standard$$
the corresponding worldsheet orbifold will be referred to as a ``standard''
worldsheet orbifold.  The complementary case, presumably more interesting,
where the worldsheet group $\wsgroup$ mixed non-trivially with a target group
action, is referred to as an ``exotic'' worldsheet orbifold.

At genus $g$, the partition function of a worldsheet orbifold CFT receives
contributions from surfaces with $h$ handles, $b$ boundaries and $c$
crosscaps, with $\frac{1}{2}b+\frac{1}{2}c+h=g$.  On each particular surface
$\Sigma_g$, the partition function contains the sum over all possible
monodromies on $\Sigma_g$:
$$Z_{\Sigma_g}(m)=\frac{1}{\left| G\right|^g}\sum_{\alpha :\pi_1(\Sigma_g)
\rightarrow G}Z_{\Sigma_g}(\alpha ;m),\eqn\eemonod$$
where $m$ are the moduli, $\pi _1(\Sigma _g)$ is the (orbifold) fundamental
group of $\Sigma _g$, and $Z_{\Sigma _g}(\alpha ;m)$ denotes the amplitude
calculated with the particular set of monodromies $\alpha$.  For exotic
worldsheet orbifolds, the representations of $\pi _1(\Sigma _g)$ to be summed
over, are constrained analogously as in \tobetriangle .  $\pi _1(\Sigma _g)$
is a $\ztwo$ extension of the fundamental group of the double of $\Sigma _g$.
Hence, there is a natural projection of $\pi _1(\Sigma _g)$ to $\wsgroup$, and
the allowed monodromies complete the following diagram,
$$\pi _1(\Sigma _g)\rightarrow \wsgroup \leftarrow G,\eqn\eecomple$$
to a commutative triangle.  For example, the amplitude on the cylinder reads
$$Z_{\rm C}(t)=\frac{1}{\left|G\right|}\sum_{g_1,g_2,h}Z_{\rm C}(g_1,g_2,h;t),
\eqn\eecyl$$
where the monodromies are of the form $g_i=\tilde{g}_i\times \Omega$,
$h=\tilde{h}\times 1$, as elements of $G\subset \widetilde G\times\wsgroup$,
and satisfy
$$g_1^2=g_2^2=1,\qquad[g_i,h]=1.\eqn\eesquare$$

Much information about any theory is encoded in its one-loop amplitudes.  In
string theory, one-loop \foot{in the string coupling constant} diagrams
correspond to genus-one topologies of the worldsheet; in unoriented open and
closed string theory, they are given by the torus, Klein bottle, cylinder, and
M\"obius strip.  The amplitudes can be computed in two different pictures
[\bc].  The loop-channel picture corresponds to open and closed strings
comprising loops of length $t$ (with the width of the strings properly
normalized).  In this picture, the amplitudes can be calculated conveniently
as traces over corresponding Hilbert spaces of closed and open strings.  The
tree-channel picture corresponds to a cylinder of length $\tilde t$
created from and annihilated to the vacuum via boundaries and crosscaps; the
moduli $t$ and $\tilde t$ of the two channels are related by $t=1/(2\tilde t)$
for the Klein bottle and the M\"obius strip, and by $t=2/\tilde t$ for the
cylinder.  It is well-known [\bc] that the boundary and crosscap conditions on
the fields can be translated into the quantum mechanical language by
constructing the corresponding boundary and crosscap states $\ket B,\ket C$.
This construction gives a simple recipe for calculating amplitudes in the tree
channel.  In the tree channel, the amplitude corresponds to the creation of a
closed string from the vacuum by $\bra B$ or $\bra C$, subsequent free closed
string propagation, and final annihilation into the vacuum by either $\ket B$
or $\ket C$.  Comparing these two ways of computing the one-loop amplitudes we
get a set of constraints:
$$\eqalign{{\rm Tr}_{\rm open}\left( e^{-H_ot}\right)&=\bra B
e^{-H_c\widetilde t}\ket B,\cr
{\rm Tr}_{\rm open}\left( \Omega e^{-H_ot}\right)&=\sfrac{1}{2}\left\{
\left\langle B\right| e^{-H_c\widetilde t}\left| C\right\rangle +\left\langle
C\right| e^{-H_c\widetilde t}\left| B\right\rangle \right\},\cr
{\rm Tr}_{\rm closed}\left( \Omega e^{-H_ct}\right)&=\Bra Ce^{-H_c
\widetilde t} \Ket C,\cr}\eqn\ZMS $$
analogous to the requirements of modular invariance in closed CFT.  The factor
of one half in the middle equation of \ZMS\ is explained by observing that the
$\ztwo$ symmetry that interchanges the two boundaries of the cylinder, or the
two crosscaps of the Klein bottle, is to be divided out as a part of the gauge
group in the full-fledged string theory, but not in CFT that we are
considering here.

The one-loop conditions \ZMS\ pose stringent consistency restrictions on the
theory.  If we calculate amplitudes for a given model, say, in the loop
channel, we must check whether corresponding boundary and crosscap states
exist such that \ZMS\ be valid.  Moreover, any relative normalization of the
boundary state against the crosscap state, motivated e.g.\ from the BRST
invariance in full-fledged string theory or from modular geometry in CFT,
fixes the normalization of the loop-channel expressions.  This normalization
then self-consistently determines the Chan-Paton degeneration of the open
sector of the string spectrum.  This is an outline of how the Chan-Paton
symmetry in open strings is controlled by modular geometry.
%
\section{A Thickening of the Open String}

In worldsheet orbifold models, left and right movers are coupled to each other
through boundaries and/or non-orientability of the worldsheet.  To find a
correspondence of this coupling between left and right movers in the CS gauge
theory, we have to identify how CFT with both left and right sectors enters
CS gauge theory.  In the case of CFTs on closed oriented surfaces, an answer
to this question was conjectured by Witten [\Wgrav] and further developed by
Moore et al.\ [\MSZoo,\EMSS], Kogan, and Carlip [\Kogan,\CarlipKogan] (see
also [\HVer]).  Their results can be simply summarized as follows.

Let us quantize the theory canonically on $C\times\BR$, with $C$ a cylinder
[\EMSS].  Working in the axial gauge $A_0=0$, we must first satisfy the
constraint that requires the space-like part of the curvature to be zero,
$\widetilde F=0$.  This is easily solved to give (the tildes over
$\widetilde d$ and $\widetilde A$ denote the space-like parts of $d$ and $A$):
$$\eqalign{\widetilde A&=-\widetilde d\widetilde U\widetilde U^{-1},\cr
\widetilde U&=U\exp{\left( i\frac{\lambda }{k}\phi\right) },\cr}\eqn\eespace
$$
where $U$ is a single-valued map from $C$ to $\CG$, and $\lambda$ measures the
holonomy around the non-contractible loop on the cylinder.  Inserting this
solution into the Lagrangian \action , we can reduce it to an effective
Lagrangian for $U$ and $\lambda$:
$$\eqalign{S(U,\lambda )&=\frac{k}{4\pi }\int _{\p C\times\BR}{\rm Tr}
\left( U^{-1}\p _{\phi}UU^{-1}\p _tU\right)\,d\phi\,dt + \frac{k}{12\pi }
\int _{C\times\BR} {\rm Tr} \left( U^{-1}dU\right) ^3 \cr
&{}+\frac{1}{2\pi }\int _{\p C\times\BR}{\rm Tr}\lambda (t)\left(
U^{-1}\p _tU\right)\,d\phi\,dt.\cr}\eqn\eeeffect
$$
The Hilbert space $\CH_C$ resulting from the quantization of this phase space
has the structure of
$$\CH=\bigoplus_{\lambda}\,[\phi_{\lambda}]\otimes\overline{[
\phi_{\lambda}]},\eqn\eehilb$$
where $\lambda$ now belongs to the set of integrable representations of the
Kac-Moody group $\widehat\CG$, and $[\phi _{\lambda}]$ denote the
representations.  This Hilbert space exactly corresponds [\EMSS] to the
Hilbert space of the WZW model with $\widehat\CG$ as its Kac-Moody
symmetry group, and with a diagonal modular invariant.  Within this
correspondence, gauge invariant degrees of freedom living at one component of
$\p C$ correspond to the left movers, while the second component of $\p C$
yields the right movers of the WZW CFT.  Thus, the cylinder $C\equiv
S^1\times[0,1]$ is the manifold that represents the thickening of the closed
string in Chern-Simons theory, and similarly, $\Sigma\times[0,1]$ is the
three-dimensional thickening of closed oriented surface $\Sigma$.

Now we will look for an analogous 3D setting for open strings.  In the
previous subsection we have seen in an outline how open strings emerge in
twisted sectors of worldsheet orbifold models.  Now I will argue that the
orbifold construction extends also to the 3D CS theory:  We will see that a
natural thickening of the open string is a two-dimensional $\ztwo$ orbifold
with a boundary; I will also construct the thickened version of surfaces with
boundaries and/or crosscaps, as particular three-dimensional $\ztwo$
orbifolds.  The final check of the proposed correspondence then comes from the
fact that it reproduces the known structure of CFT on surfaces with boundaries
and/or crosscaps (including such subtleties as the vacuum degeneration of the
open string spectrum).

\insfig{1}{(a)~The $\ztwo$ symmetry of the cylinder $C$ that defines the
thickened open string $\CO_C$ as $\CO_C\equiv C/\ztwo$.  (b)~The thickened
open string $\CO_C$. The $\ztwo$ singular points $P_1$ and $P_2$ are the only
singular points of $\CO_C$.}{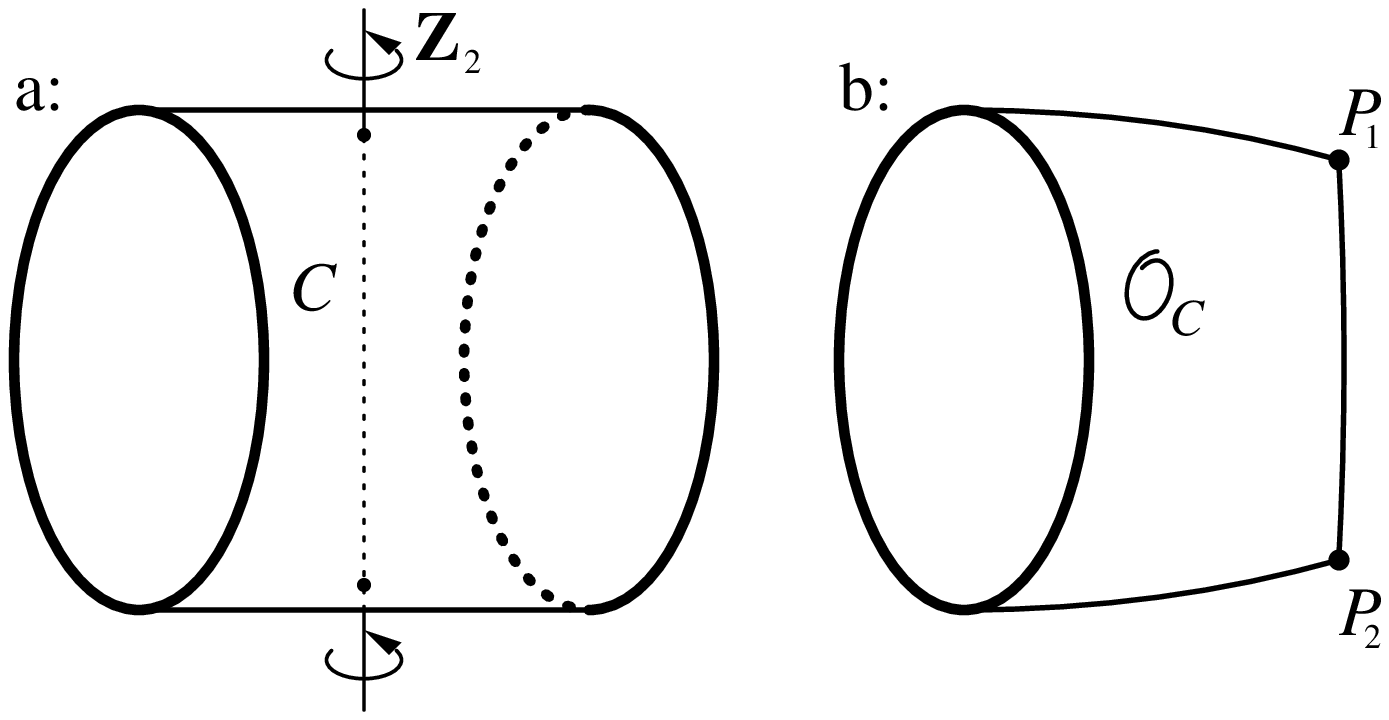}{3}

With this motivation in mind, we will proceed by studying CS gauge theory on
$\ztwo$ orbifolds.  To be a symmetry of the Chern-Simons Lagrangian \action ,
the orbifold group $\ztwo$ must act on 3D ``spacetimes'' by
orientation-preserving diffeomorphisms.  A particularly important class of
such actions are products of $\BR$ (with the trivial $\ztwo$ action) and
a 2D manifold $\Sigma$ with an orientation-preserving involution.  In
particular, we may take $\Sigma=C$, the thickened closed string, and consider
the $\ztwo$ action that interchanges the boundaries of $C$ as in
figure~(1.a).  The resulting 2D orbifold, denoted by $\CO_C$ throughout this
paper, is the proposed thickened version of the open string (cf.\
figure~(2)).

\insfig{2}{The correspondence between the orbifold $\CO_C$ and the open
string $\CO_S$.  The singular points of $\CO_C$ correspond to the boundary
points of the open string, while the boundary of $\CO_C$ corresponds to the
interior of the open string worldsheet.}{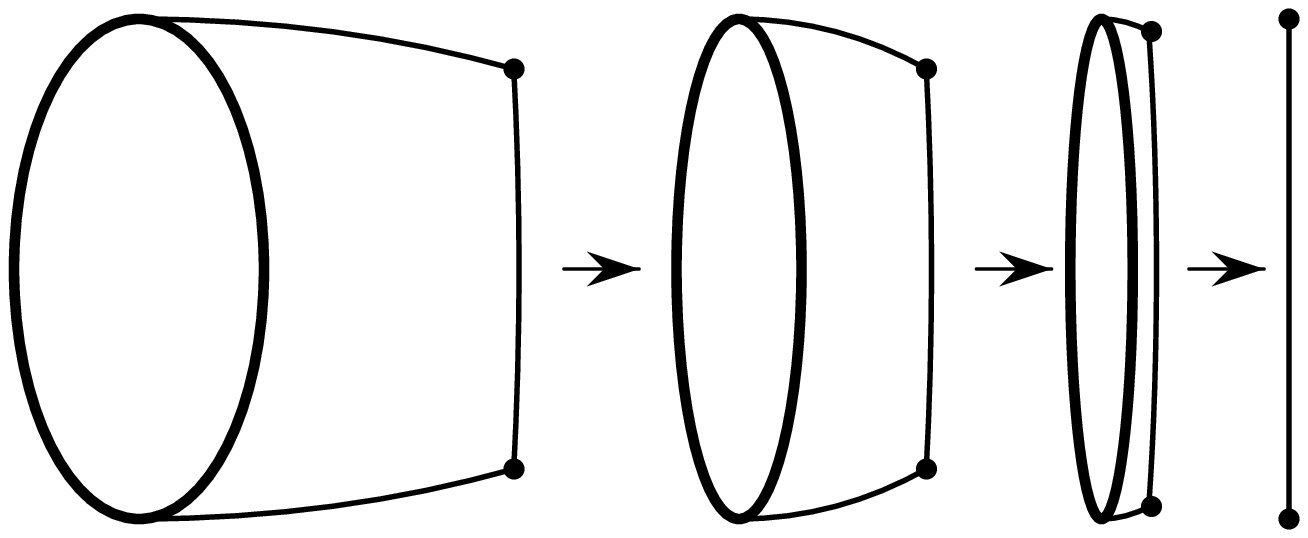}{3}

\insfig{3}{The first homotopy group of $\CO_C$, $\pi_1(\CO_C)=\homotopy
\equiv\ztwo\semi\BZ$.  The picture shows the generators $\gamma_1,\gamma_2$ of
the $\ztwo$ components in $\homotopy$, as well as their product $\gamma$ that
generates the normal subgroup $\BZ$ in $\ztwo\semi\BZ$.}{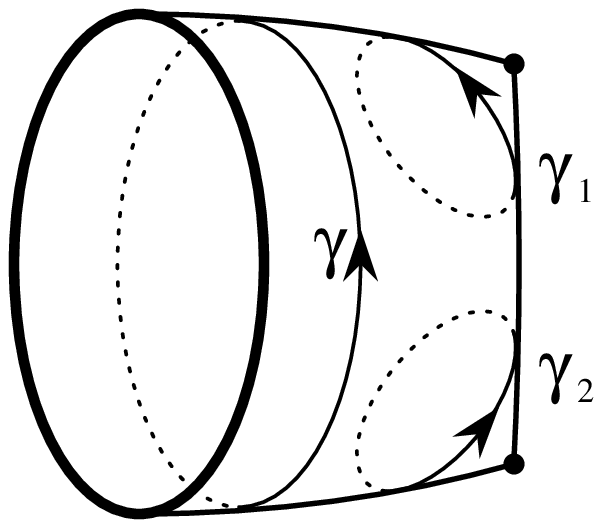}{3}

One fact that supports this correspondence is the isomorphism between the
first homotopy groups of the thickened open string $\CO_C$ and the open string
$\CO_S$, which are both isomorphic to $\homotopy$.  For the thickened open
string, the structure of the first homotopy group is indicated in figure~(3).
In particular, the generators of $\pi_1(\CO_S)$ that correspond to the
``boundary twists'' on the open string (cf.\ \twistplus, \twistminus) now
correspond to the non-contractible circles wrapped around the singular points
of $\CO_C$.  (Actually, the ``correspondence'' of figure~(2) is a homotopy
equivalence in the corresponding category of orbifolds; cf.\ Appendix A.)

We have seen that any closed oriented worldsheet $\Sigma$ of closed string
theory can be naturally thickened to a three-manifold $M=\Sigma\times[0,1]$.
As for surfaces of worldsheet orbifold models, i.e.\ surfaces with boundaries
and/or crosscaps, we can construct their natural thickening as follows.  Let
$\Sigma$ be a surface with boundaries and/or crosscaps, and $\bar\Sigma$ its
oriented double with empty boundary.  Denote by $I$ the defining involution on
$\bar\Sigma$, i.e.\ $\Sigma=\bar\Sigma/I$.  The corresponding thickening of
$\Sigma$ is then
$$\CO_\Sigma=(\bar\Sigma\times[0,1])/I,\eqn\eethick$$
where $I$ acts on $t\in [0,1]$ via $t\rightarrow 1-t$.  $\CO_\Sigma$ is an
orbifold with boundary, $\p\CO_\Sigma$ being isomorphic to one component of
$\bar\Sigma$.  Two examples of such thickened open string diagrams are shown
in figure~(4).

\insfig{4}{The thickened versions of (a) the annulus, and (b) the M\"obius
strip diagrams. The thick lines represent the singular loci of the
orbifolds;  the shaded two-dimensional sections are isomorphic to the
thickened open string $\CO_C$.}{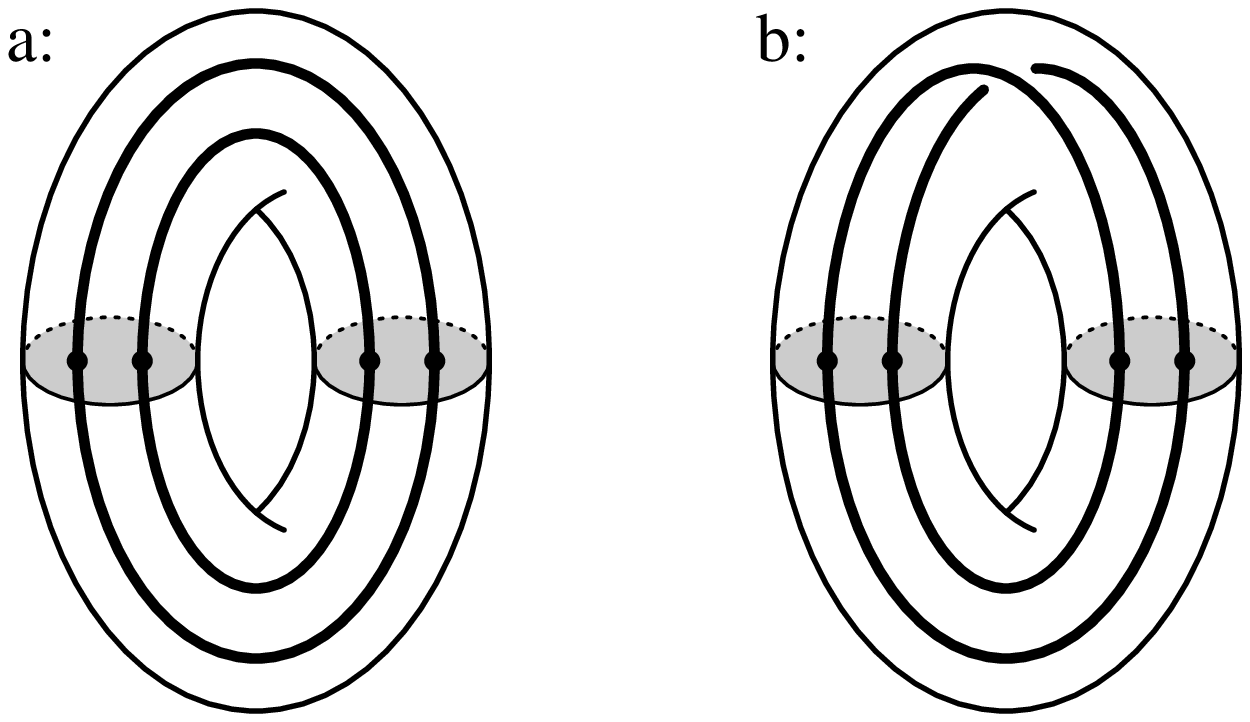}{3}

At the quantum level, there is a correspondence between the partition function
of the two dimensional WZW model on a closed oriented surface $\Sigma$, and
the (transition) amplitude of the CS gauge theory on $M$, summed up over the
natural basis of $\CH_\Sigma$:
$$Z_\Sigma=\sum h_{ij}\Psi_i\otimes\overline{\Psi}_j\quad\in\quad\CH_\Sigma
\otimes\overline\CH_\Sigma.\eqn\Zpsipsi$$
The aforementioned correspondence between 2D surfaces of open string theory
and their 3D orbifold thickenings leads to an open string counterpart of
\Zpsipsi ,
$$
Z_\Sigma=\sum a_i\Psi_i\quad\in\quad\CH_{\bar\Sigma}.\eqn\eeopen$$
Here $\Sigma$ is a surface with at least one boundary or crosscap, and
$\bar\Sigma$ denotes its double.

This picture allows us to make, already at this stage, some preliminary
conjectures about the relation between 3D CS gauge theory on orbifolds and
2D CFT of open strings.  The closed string, which is topologically a circle,
can be obtained from its thickening if the boundaries of the thickening
approach each other.  In the case of open strings, the analogous procedure of
shrinking the thickening $\CO_C$ to the open string is shown in figure~(2).
The boundary points of the string correspond to the two points in the
singular locus of the thickening.  We thus expect that the structure of
Chan-Paton factors is related to the geometry of the singular locus.  On the
other hand, the bulk degreees of freedom on the worldsheet of the open string
are expected to correspond to gauge-invariant degrees of freedom at the
boundary of the thickening.  These expectations will be confirmed below.

%
\chapterr{Quantization of CS Gauge Theory on Orbifolds}

We have seen in the previous section that the natural setting for the
Chern-Simons counterpart of 2D CFT on surfaces with boundaries and crosscaps
is the theory on three dimensional $\ztwo$ orbifolds, and we have made several
preliminary conjectures about the correspondence between these two theories.
In order to substantiate these expectations, we must quantize the Chern-Simons
theory on orbifolds and compare the outcome to the structure of 2D CFT of
worldsheet orbifolds.

As a first step towards the definition of the quantum CS gauge theory with
gauge group $\CG$ on an orbifold, we have to specify a Lagrangian for
connections on any principal $\CG$-bundle over arbitrary orbifold $\CO$.  One
is tempted to define, in analogy with string theory on orbifolds, the
Lagrangian on an orbifold $\CO$ for a given $\CG$-bundle $E$ via the
Lagrangian of CS gauge theory on the doubling $\bar\CO$ of $\CO$:
$$2S(A)=S(\bar A)\equiv\frac{k}{4\pi}\int_{\bar\CO}{\rm Tr}\left(\bar A\wedge
d\bar A+\frac{2}{3}\bar A\wedge\bar A\wedge\bar A\right).\eqn\orbaction$$
Here $\bar A$ is the pullback of the connection $A$ from $\CO$ to the
doubling bundle $\bar E$ over the doubling manifold $\bar\CO$.  The formula
is well defined at least for compact, connected, simply connected gauge
groups, $\bar E$ being in this case the trivial principal bundle over the
manifold $\bar\CO$.  Nevertheless, this definition is still incomplete, since
we have to resolve the ambiguity that has emerged because we have defined
a {\em multiple} of $S(A)$ in \orbaction , and we have to resolve this
ambiguity for any orbifold in a way compatible with factorization (see
[\DijkW] for a thorough discussion of this argument in a slightly different
context).  For general gauge groups, this requires techniques of equivariant
cohomology of classifying spaces, and I refer the reader to Appendix B, where
the general answer is presented.

In order to avoid technical complications, I will now focus on $\CG$ compact,
connected, and simply connected.  The classical phase space to be canonically
quantized on a Hamiltonian slice $\Sigma$ of a three dimensional
``spacetime,'' is given by
$$\CP(\Sigma)={\rm Hom}\left(\pi_1(\Sigma),\CG\right)\times{\rm Maps}\left(
\p\Sigma,\CG\right).\eqn\phase$$
One basic building block of this phase space is the space of possible
holonomies around a singular point.  Ignoring temporarily the overall
conjugation by $\CG$, it is given by all representations of $\ztwo$ in $\CG$,
i.e.\ the submanifold $\pointph$ in $\CG$ of those elements whose square is
one.  The phase space $\pointph$ pierces a fixed maximal torus $\CT$ in a
finite set $\CT_{(2)}$, and in turn, $\pointph$ can be recovered from this
finite set by conjugating $\CT_{(2)}$ by $\CG$.  Thus, $\pointph$ can be
decomposed into conjugacy classes $\omega e^\lambda\omega^{-1}$ classified by
$e^\lambda\in\CT_{(2)}$.

Let us specialize for simplicity to $\Sigma=\CO_D$, the disk with one singular
point inside, and define the Chern-Simons Lagrangian on the unconstrained
phase space, restricting ourselves to the holonomies that are conjugated to a
particular element $e^\lambda$ of $\CT_{(2)}$.  The general case can be
treated similarly.  Respecting all the required symmetries, we get the
Lagrangian that combines the usual Chern-Simons Lagrangian with the coadjoint
orbit Lagrangian for the holonomies around the singular point:
$$\eqalign{S(A,\omega)&=\frac{k}{8\pi}\int_{\bar\CO_D}{\rm Tr}\left(\bar A
\wedge d\bar A+\sfrac{2}{3}\bar A\wedge\bar A\wedge\bar A\right)\cr
&{}+\int dt\,{\rm Tr}\left(\lambda\omega^{-1}(t)\left(\p_t+A_0\right)\omega(t)
\right),\cr}\eqn\aa$$
with the notation of \orbaction , and $\omega$ parametrizing the component of
$\pointph$ consisting of the elements conjugated to $e^\lambda$.  This
Lagrangian is anomalous unless $\lambda$ is a weight [\EMSS].  This condition
poses restrictions on possible values of $k$ (cf.\ [\EMSS]); I will limit the
discussion henceforth to the non-anomalous $k$'s.  Consequently, quantization
of the corresponding effective Lagrangian on the constrained phase space $\CP$
leads to the Hilbert space consisting of irreducible representations
$[\phi_\lambda]$ of the loop group $\CL\CG$, with $\lambda$ being from
$\CT_{(2)}$.  In this section I will only discuss the simplest case of
$\CG=\su 2$, unless stated otherwise.

On our favorite orbifold $\CO_C$, the reduced phase space $\CP$ consists of
the product of two copies of $\pointph$, times the space of gauge invariant
degrees of freedom that survive at the boundary.  In the case of $\CG =\su 2$,
the set of holonomies allowed around the singular points reduces to two
points, corresponding to the representations of spin $0$ and $k/2$ at level
$k$, which is then necessarily even.  Quantization of the corresponding phase
space leads to the Hilbert space
$$\CH=2\left\{ [\phi_0]\oplus[\phi_{k/2}]\right\}.\eqn\opspectrum$$
According to the correspondence between CS gauge theory and two dimensional
CFT of open strings, we expect this space to represent the Hilbert space of
open string states of a worldsheet orbifold of the $\su 2$ WZW model.  I will
demonstrate that this is indeed the case in \S{4.1}, where I identify
explicitly the CFT that corresponds to \opspectrum .

%

\section{The Singular Locus as a Link of Wilson Lines}

We have seen in \aa\ that the points in which the singular locus pierces a
chosen Hamiltonian slice effectively behave as sources of curvature for the
Chern-Simons gauge field.  More precisely, the form of the Lagrangian \aa\
indicates that the singular locus is effectively equivalent in the quantum
theory to a sum of Wilson lines in some particular representations of the
gauge group.  This simple but important fact allows us to reduce the theory on
orbifolds to a theory on manifolds, trading the singular locus for a link of
Wilson lines.

Another argument that will allow us to see the equivalence, follows closely
the reasoning of [\Wstat].  Consider a connected component of the singular
locus in a 3D orbifold $M$, and denote it by $\ell$.  It can be surrounded by
a 2D torus $T$, which divides $M$ into two disconnected parts, i.e.\ a solid
torus with $\ell$ inside it, and the remnant.  Whatever happens inside the
solid torus, defines a vector from $\CH_T$.  A natural basis in $\CH_T$ is
given by functional integrals over the solid torus with all the allowed Wilson
lines replacing $\ell$.  The vector that describes the functional integral
with the component $\ell$ of the singular locus inside the solid torus can be
expanded in this basis,
$$\ell=\sum_{R_i}c_{R_i}W_{R_i}(\ell),\eqn\locuscorresp$$
where $c_{R_i}$ is a set of complex numbers.  Effectively, all information
about the presence of orbifold singularities is now stored in these numbers.

We have just argued that any connected component $\ell$ of the singular locus
on an orbifold $\CO$ can be represented as a sum over Wilson lines with the
topology of $\ell$.  As a result of this equivalence, the theory on orbifolds
is reduced to the ``parent'' theory on manifolds,
\foot{In some more complicated cases, discussed in \S{4.2} and \S{5} below,
the gauge group of the ``parent'' theory may differ from $\CG$ by a discrete
factor.}
as follows.  Using \locuscorresp , the correlation function of an arbitrary
collection of physical observables $\Phi$ on $\CO$ can be calculated as a sum
over more complicated correlation functions on the underlying manifold
$X_\CO$,
\foot{$X_\CO$ is topologically the same as $\CO$ but with orbifold
singularities smoothed out.  We are safe here, at least if $X_\CO$ is a
topological manifold, because every three dimensional topological manifold
admits exactly one differentiable structure.}
with the singular locus traded for a link of specific Wilson lines:
$$\eqalign{{\vev\Phi}_\CO&=\vev{\Phi\prod_\alpha\left(\sum_{R_i}c_{R_i}W_{R_i}
(\ell_\alpha)\right)}_{X_\CO}\cr
&\qquad\equiv\sum_{R_i^{(1)},\ldots R_j^{(s)}}c_{R_i^{(1)}}\ldots c_{R_i^{(s)}}
\;\vev{\Phi\ W_{R_i^{(1)}}(\ell_1)\ldots W_{R_i^{(s)}}(\ell_s)}_{X_\CO}.\cr}
\eqn\correlcorresp$$
(Here $\alpha=1,\ldots s$ counts the connected components $\ell_\alpha$ of the
singular locus, and $X_\CO$ is the underlying manifold of $\CO$.)  In
particular, the partition function of the CS gauge theory on the orbifold is
equivalent to a correlation function of the usual CS gauge theory on on the
underlying manifold:
$$\eqalign{Z(\CO)&=\vev{\prod_\alpha\left(\sum_{R_i}c_{R_i}W_{R_i}(\ell_\alpha)
\right)}_{X_\CO}\cr
&\qquad\equiv\sum_{R_i^{(1)},\ldots R_j^{(s)}}c_{R_i^{(1)}}\ldots c_{R_i^{(s)}}
\;\vev{W_{R_i^{(1)}}(\ell_1)\ldots W_{R_i^{(s)}}(\ell_s)}_{X_\CO}.\cr}
\eqn\partition$$
These two formulas represent one of the central points of this paper.  They
relate the correlation function in a theory on orbifolds (which we {\it a
priori} do not know how to calculate) to a sum of more complicated correlation
functions in the simpler theory on manifolds (which we do know how to
calculate).

To establish the correspondence between the theory on orbifolds and the theory
on the underlying manifolds, it now only remains to determine the $c_{R_i}$'s
of \locuscorresp .  To this aim let us consider the theory on an orbifold
which is topologically a solid torus, with the singular locus isomorphic to
the generator of the fundamental group.  This functional integral determines a
state from the Hilbert space on the torus.  We can measure this state by the
following procedure.  Let us take another copy of the solid torus, now with an
arbitrary Wilson line $W_R(b)$ replacing $\ell$, with $b\sim\ell$
topologically, and glue these two solid tori together, so as to obtain
$S^2\times S^1$.   The functional integral of the resulting object is easily
calculable as a trace over the physical Hilbert space of the twice punctured
sphere.  On the other hand, the same amplitude is equal to the inner product
of the states that result from functional integrals over the solid tori before
gluing.  This leads to the following formula, which allows one to determine
$c_{R_i}$:
$$Z(S^2\times S^1,R,\ell)=\sum_{R_i}c_{R_i}\left(v_{R_i},v_R\right).
\eqn\eedeter$$
Here $(\ ,\ )$ denotes the inner product in $\CH_T$, and $R_i$ are the
representations carried by the singular locus.  This completes the arguments
on the equivalence \locuscorresp\ between the singular locus and a sum of
Wilson lines with the same topology.  As a sample application of this
equivalence, note that each component of the singular locus in the $\su 2$
theory is equivalent to $W_0(C)+W_{k/2}(C)$.
%

\section{Framing of the Singular Locus}

It is known that in quantum CS gauge theory, Wilson lines need framing.  In
particular, the singular locus, being equivalent to a sum of links of Wilson
lines, may need framing.  One may thus wonder how the statements of the
previous paragraph interfere with this additional structure needed for a well
defined quantum theory.

First of all, note that the singular locus of any orbifold required for the
correspondence to 2D CFT can be canonically framed.  We have seen in \S{2.2}
that he orbifolds representing the thickening of an open string surface are of
the form
$$\CO=(\Sigma\times[0,1])/\ztwo.\eqn\eesotired$$
Such an orbifold can indeed be retracted uniquely (up to homotopy) to the two
dimensional surface $\Sigma/\ztwo$.  Thus, we can pick an arbitrary imbedding
of the two dimensional surface $\Sigma/\ztwo$ into $\CO$, with boundaries
mapped to the singular locus of $\CO$.  This retraction gives a unique and
natural framing to the singular locus, simply by demanding that the vectors
that frame the singular locus are tangent to the image of $\Sigma/\ztwo$, and
point inward.

Since the canonical framing of the singular locus always exists (and is
unique) for the orbifolds that represent the thickened open string surfaces,
we need not worry about framing in the applications to 2D CFT on surfaces with
boundaries and/or crosscaps; we would still need something more, however, were
we interested in the full-fledged CS gauge theory on general $\ztwo$
orbifolds.  Results of [\Aframings], which indicate that there might be a
preferred way how to frame a three-manifold, are particularly interesting in
this context.  Alternatively, we could restrict ourselves to those models that
do not require framing of the singular locus, i.e.\ do not require framing of
the particular set of Wilson lines that effectively represent the singular
locus in the correlation functions according to \locuscorresp .  This
restriction would impose an additional condition on the CS coupling constant
$k$.  For example, in the case of $G=\su 2$ that we have been focusing on in
this section, the singular locus carries the representations with spin $0$ or
$\frac{k}{2}$.  If the framing of a Wilson line $W_R(C)$ is shifted by a
$t$-fold twist, the corresponding state is multiplied by $e^{2\pi ih_Rt}$,
where $h_R$ is the conformal weight of the primary field corresponding to
$R$.  Conformal weights of the primaries $\phi_j$ of the $\su 2$ WZW model
are
$$h_j=\frac{j(j+1)}{k+2};\eqn\weight$$
hence, the conformal weight of the non-trivial primary $\phi_{k/2}$ carried by
the singular locus of the $\su 2$ theory equals $h_{k/2}=\frac{k}{4}$.
Insisting on the integrality of the conformal weights of the primaries that
correspond to the singular locus, we get the restriction $k=0\ ({\rm mod}\ 4)$
on the coupling constant of the CS gauge theory on orbifolds.
%

\section{Skein Theory for the Singular Locus}

One of the most appealing and important properties of Chern-Simons correlation
functions of Wilson lines is their calculability by (un)braiding the Wilson
lines using skein theory.  To a given two dimensional surface $\Sigma$ with
$p$ punctures and representations $R_i$, $i=1,\ldots p$ inserted in them, CS
gauge theory assigns the Hilbert space $\CH_{\Sigma ,R_i}$ of physical states,
which is $n$-dimensional.  The skein relations are linear dependence
relations, satisfied by any set of $n+1$ vectors of this vector space.

This braiding procedure plays an interesting role in the comparison with CFTs
of worldsheet orbifolds.  Indeed, the only difference between the thickened
cylinder (figure~(4.a)) and the thickened M\"obius strip (figure~(4.b)) is in
braiding of the singular locus.  More explicitly, the functional integral over
these two topologies, with the particular labeling of the Wilson lines, gives
the partition functions of the associated 2D CFT on the worldsheet of the
topology of the cylinder and M\"obius strip respectively, $\Psi_{\rm C}$ and
$\Psi_{\rm MS}$, which are elements of the Hilbert space of the CS gauge
theory on the torus.  Using the results of \S{2} and \S{3.2}, we have
$$\Psi_{\rm C}=\sum_{R,R'\in\{0,\frac{k}{2}\}}\Vev{W_R(\ell)W_{R'}(\ell')}_{
{\rm solid}\;{\rm torus}},\eqn\psione$$
and
$$\Psi_{\rm MS}=\sum_{R\in\{0,\frac{k}{2}\}}\Vev{W_R(\widetilde\ell)}_{{\rm
solid}\; {\rm torus}}.\eqn\psitwo$$
Here $\ell ,\ell',\widetilde\ell$ denote components of the singular loci
as shown in figure~(4).  The only difference between the two orbifolds can be
localized within a small two-sphere, pierced four times by the singular
locus.  Cutting out the ball surrounded by this two-sphere, we get an orbifold
$\CO$ whose boundary $\p M$ is isomorphic to the disconnected sum of the torus
and the four punctured sphere.  Then we can compute the $\Psi_{\rm C}$ and
$\Psi_{\rm MS}$ of \psione, \psitwo\ as inner products in the Hilbert space
of the four-times punctured sphere $\CH_{S^2}$:
$$\Psi_{\rm C}=\left(u,v\right),\qquad\Psi_{\rm MS}=\left(u,v'\right),
\eqn\relation$$
where $u\in\CH_{S^2}$ represents the functional integral over $\CO$, and
$v,v'$ are the functional integrals over the three-balls with Wilson lines as
shown in figure~(5.b).

\insfig{5}{Skein relations of the singular locus. $R$ and $R'$ are the $\su 2$
representations that can be carried by the singular locus, i.e.\ their spins
are either $0$ or $\frac{k}{2}$, and $k$ is assumed to satisfy $k=0\ ({\rm mod}
\ 4)$.}{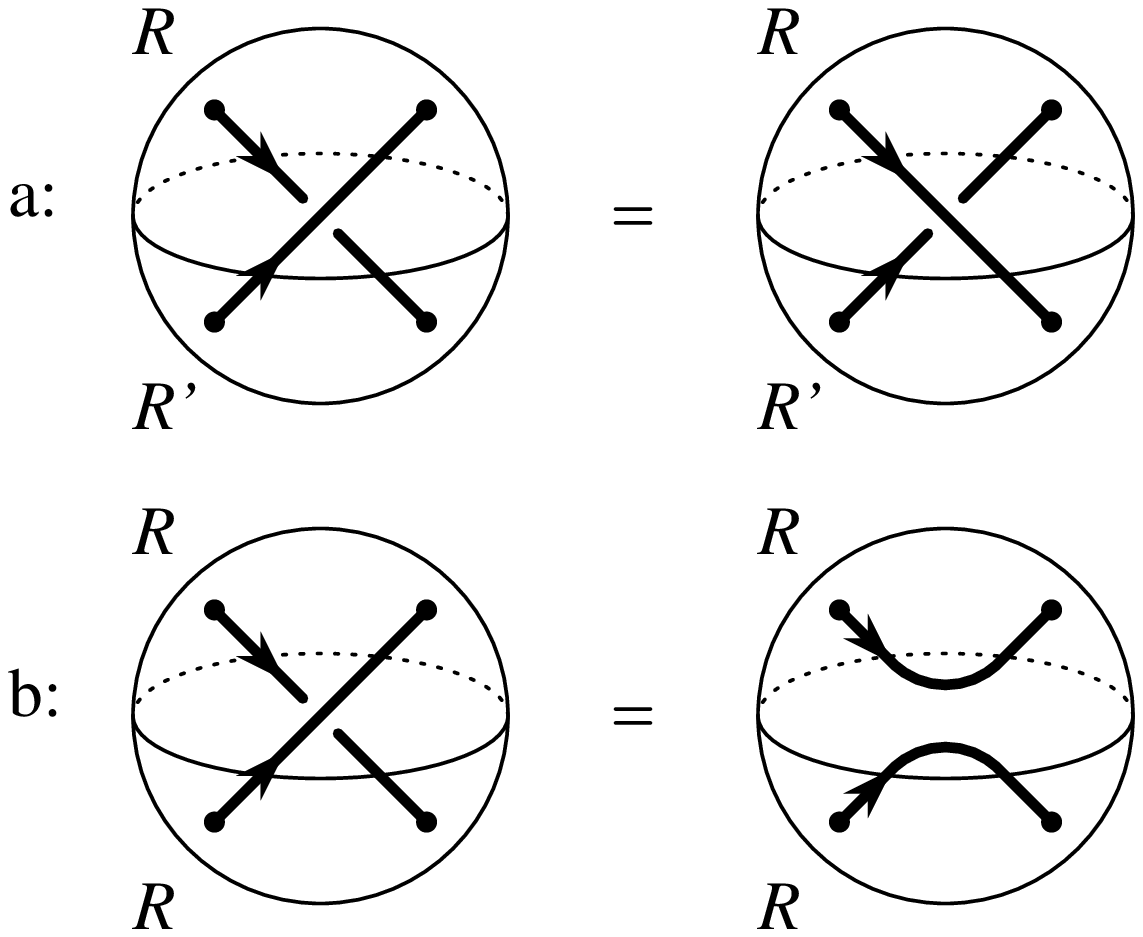}{4}

Let us now restrict ourselves to $\CG =\su 2$ with $k=0\ ({\rm mod}\ 4)$, for
which we have seen in the previous subsection that the theory is independent
of framing of the singular locus.  The singular locus is equivalent to a sum
of Wilson lines with  $R,R'\in \{0,\frac{k}{2}\}$.  For such $R,R'$, the
corresponding Hilbert space is one-dimensional, as can be easily inferred from
the fusion rules of the $\su 2$ WZW model [\CFTag]:
$$[\phi_{j_1}]\times [\phi_{j_2}]=\sum_{j=|j_1-j_2|}^{j=\min (j_1+j_2,\,k-j_1
-j_2)}[\phi_j],\qquad j_1,j_2,j\in \{ 0,\frac{1}{2},\ldots\frac{k}{2}\} .
\eqn\fusionrules$$
Thus, any two states of the physical Hilbert space are linearly dependent.  In
particular, with our restriction on $k$, the vectors given by the functional
integrals over the three dimensional balls with the Wilson lines as in
figure~(5.a) are equal to each other, the same being true of the amplitudes in
figure~(5.b).  As a consequence, the action of the braid group on the singular
points of $\CO_C$ reduces to the action of the permutation group, and any
multiple twist can be trivially unbraided.


\section{Observables}

Since the CS gauge theory on orbifolds can be effectively reduced to a CS
gauge theory on manifolds, observables on orbifolds are of precisely the same
structure as those on manifolds (with an obvious orbifold-like projection
included).  In particular, the Wilson lines indicated in figure~(6) are
natural candidates for observables.  Note that, using the equivalence
\correlcorresp\ of the singular locus and a link of Wilson lines, we can
interpret the observable $V$ of figure~(6.b) as a trivalent graph, with the
remaining two legs corresponding to the singular locus that pierces the
hamiltonian slice at the endpoint of $V$.

\insfig{6}{The topology of observables in a hamiltonian slice of the thickened
open string. The Wilson line denoted by the dashed line corresponds to (a)~the
thickened closed string vertex operator, (b)~the thickened open string vertex
operator.}{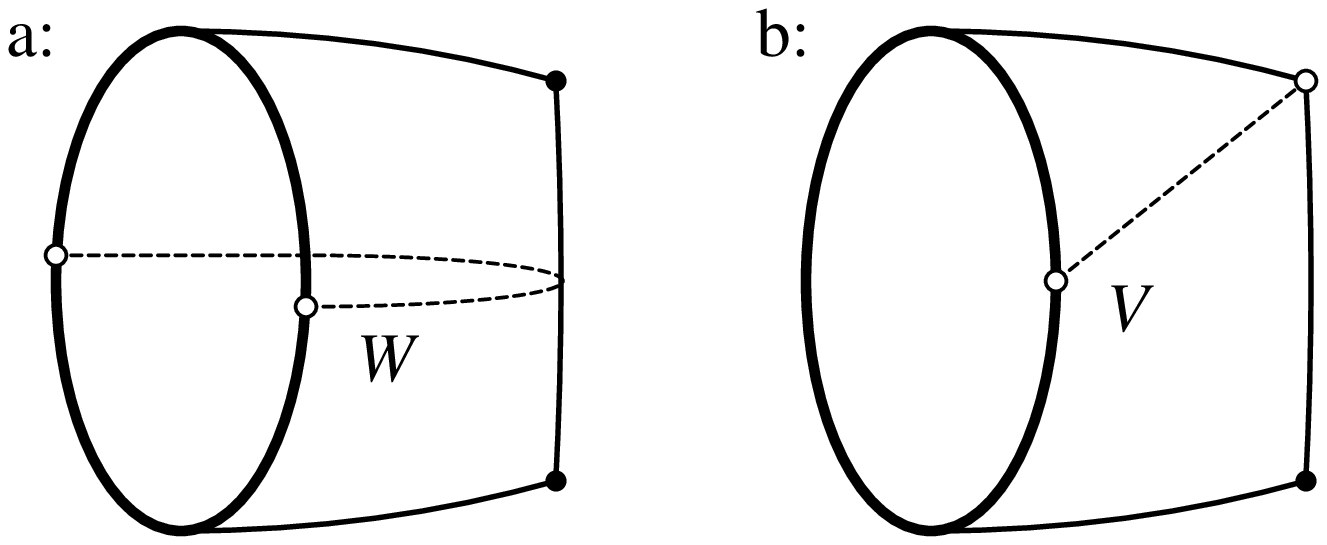}{3}

In the closed string case, Chern-Simons counterparts of 2D vertex operators
were identified by Kogan and Carlip [\Kogan,\CarlipKogan] with the Wilson
lines going from one boundary of a thickened worldsheet $\Sigma \times [0,1]$
to the other (possibly with some gauge invariant quantities attached at their
ends).  Indeed, these Wilson lines transform under a gauge transformation $g$
as the product of one left-moving and one right-moving Kac-Moody primary of
the WZW CFT model:
$$W\rightarrow g(x)\cdot W\cdot g(y)^{-1},\eqn\eeprim$$
where $x$ and $y$ are the endpoints of the Wilson line.

\insfig{7}{The three dimensional orbifold that represents the thickened
version of the open string interaction.}{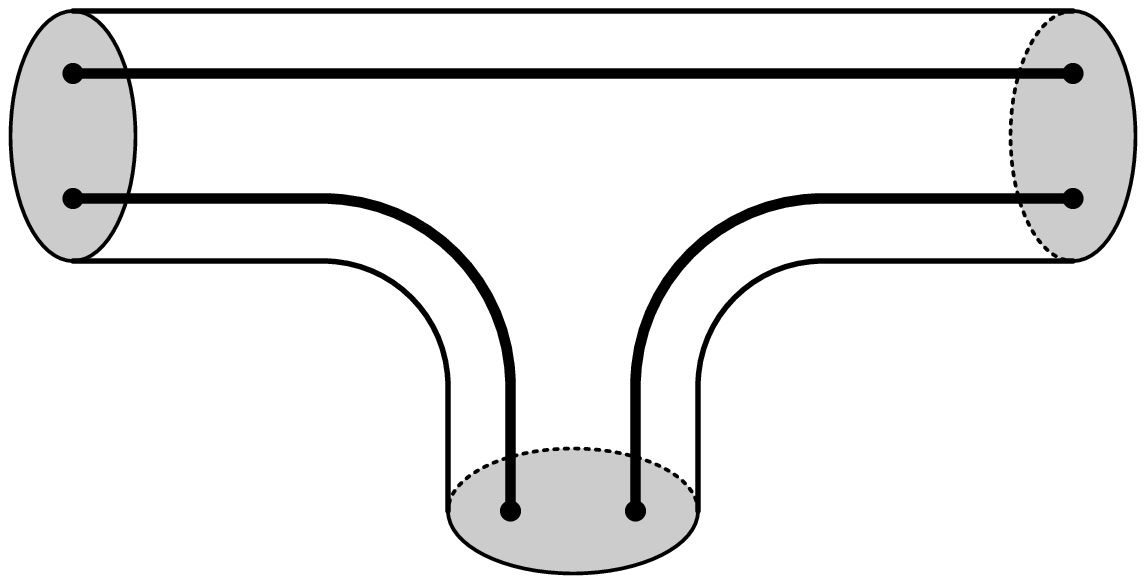}{3}

The thickened version of the open string three vertex is the ``solid pant''
orbifold of figure~(7).  It represents an interpolation between three
thickened open strings $\CO_C$.  (To get rid of the $\ztwo$ odd states in the
open string sector, we should sum over all possible permutations of the ends
of single components of the singular locus.)  With the use of \partition , the
singular locus can be traded for a sum of Wilson lines, hence the transition
amplitudes between two-dimensional orbifolds are in principle computable as a
non-trivial scattering problem on the underlying manifolds!  This is an
interesting Chern-Simons incarnation of the old idea that Chan-Paton charges
at the ends of strings represent dynamical particles (quarks of the old dual
models).  The necessity of summation over all permutations in this scattering
process resembles the analogous statements in the gravitational scattering in
$2+1$ dimensions [\Carlip].

\insfig{8}{The thickened version of the open string/closed string
interaction.}{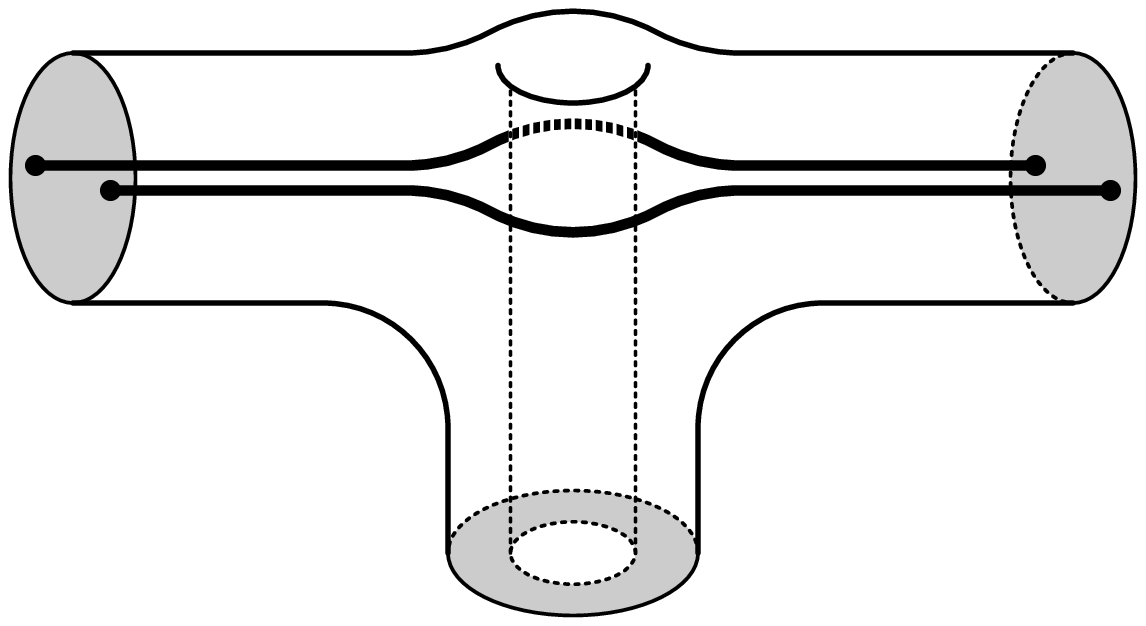}{3}

As for the closed string/open string interaction, its thickened version is
shown in figure~(8).  At the level of fundamental groups, the interaction is
equivalent to an action of the fundamental group of the thickened closed
string $C$ on the fundamental group of the thickened open
string $\CO_C$.  To see this more explicitly, let us denote by $\Gamma$ the
generator of $\pi_1(C)=\BZ$, and by $\gamma_1,\gamma_2$ (respectively
$\gamma'_1,\gamma'_2$) generators of the two $\ztwo$ components of $\pi_1
(\CO_C)=\homotopy$ before (respectively after) the interaction.  Then the
interaction acts on $\pi_1(\CO_C)$ as follows:
$$(\gamma_1,\gamma_2)\rightarrow(\gamma'_1,\gamma'_2)=(\gamma_1,\Gamma
\gamma_2\Gamma^{-1}).\eqn\homotaction$$
With this picture of thickened string interactions now at hand, it is not too
complicated to see that the infinitesimal versions of the interactions shown
in figures~(7) and~(8) are the Wilson lines of figure~(6).  In particular, the
open-string emission from an open string is represented in CS gauge theory by
a trivalent-graph observable.

%
\chapterr{The Chern-Simons/CFT Correspondence and First Examples}

The relation between CS gauge theory on orbifolds and CFT of worldsheet
orbifolds leads to a surprisingly simple prescription for the open string
spectrum of a 2D worldsheet orbifold CFT, including its Chan-Paton
degeneration, once the corresponding Chern-Simons gauge group is known.

The prescription can be briefly sketched as follows.  Given a worldsheet
orbifold CFT, consider first the restriction of the theory to closed oriented
surfaces.  Next, identify the Chern-Simons gauge group that is
associated with this ``parent'' CFT on closed oriented surfaces, in the
classification of [\MSZoo].  To get from this ``parent'' Chern-Simons theory
on manifolds to a theory on $\ztwo$ orbifolds, we must identify which class
of holonomies is allowed around the singular locus.  To this goal, in analogy
with the quantization of (3.3) in \S{3} above, we identify the primaries that
correspond to those elements of the gauge group that square to one.  Let us
denote the set of these primaries by $\CR=\{\phi_r,\,r=1,\ldots N\}$.  These
primaries are examples of the boundary states in the 2D model on surfaces with
boundaries and/or crosscaps, described first by Cardy in [\Cardy] (cf.\ the
states denoted by $\ket{\,\tilde{l}\,}$ in [\Cardy, eqn.\ (21)]).  To obtain
the spectrum of the open sector, we label each of the two singular points of
the thickened open string by a representation from $\CR$.  Fusion rules of
these representations give the bulk part of the open string spectrum, while
the structure constants of the fusion algebra determine the Chan-Paton
degeneration of the states.  The whole spectrum results from all possible
combinations of the labeling of the singular points by elements of $\CR$.

In this form, the correspondence between CS gauge theories on $\ztwo$
orbifolds and CFTs of worldsheet orbifolds clearly does not cover all possible
CFTs.  Generically, we can associate several different sets of boundary and
crosscap conditions to a given CFT on closed oriented surfaces;  yet, the
correspondence that we have just outlined associates one preferred set of
boundary and crosscap conditions to each CS gauge group (and hence, to the
corresponding CFT on closed oriented surfaces).  The question is whether the
other types of boundary conditions can also be incorporated into the scheme.
This question will be answered in the affirmative later in the paper, after we
learn more by studying several explicit examples of the correspondence between
the CS gauge theory on $\ztwo$ orbifolds and CFT on surfaces with boundaries
and/or crosscaps.  It turns out that the general scheme will require an
extension of the standard definition of CS gauge theory that will incorporate
the orbifold group $\ztwo$ into the gauge group.


\section{$\su 2$ CS Gauge Theory and Worldsheet Orbifolds}

I conjectured at the beginning of \S{3} that the Hilbert space of the $\su 2$
CS gauge theory on the thickened open string (as given by \opspectrum) should
correspond to the open string sector of a 2D CFT.  Now I will identify this
two dimensional CFT.

Inspired by \opspectrum , we are looking for a worldsheet orbifold of the
$\su 2$ WZW model that has only two primaries in the twisted (i.e.\ open) part
of the spectrum (modulo a possible Chan-Paton degeneracy), namely
$\phi_0\equiv 1$ and $\phi_{k/2}$.  Theory of open strings on group manifolds
was studied by Ishibashi in [\Ishib].  In fact, Ishibashi constructed one
particular $\su 2$-model for each $k$ even.  These models are examples of
worldsheet orbifold models as discussed in \S{2.1}.  Ishibashi starts with a
diagonal modular invariant, and takes for the projection operator in the
closed sector the parity operator acting in the obvious way on the Kac-Moody
algebra, and trivially on the basis of primary states:
$$\Omega J^a_n\Omega^{-1}=\bar J^a_n,\qquad\qquad\Omega\left|j\otimes\bar j
\right\rangle=\left|\bar j\otimes j\right\rangle.\eqn\symaction$$
This definition specifies uniquely the Klein bottle amplitude of the model,
which in the loop channel reads
$$
Z_{\rm KB}(t)=\sum_{j\in\frac{1}{2}\BZ_k}\chi_j(2it).\eqn\eeklein$$
Using the form of the $S$ matrix for the $\su 2$ WZW model:
$$
S^l_j=\sqrt{\frac{2}{k+2}}\sin\frac{(2j+1)(2l+1)\pi}{k+2},\eqn\eesmat$$
we can transform the amplitude to the tree channel,
$$Z_{\rm KB}(\tilde{t})=\sum_{j,l\in\frac{1}{2}\BZ_k}S^l_j\chi_l(2it)=
\sum_{j\in\BZ_{k/2}}\sqrt{\frac{2}{k+2}}\cot\frac{(2j+1)\pi}{2(k+2)}\chi
_j(i\tilde{t}),\eqn\eetree$$
and infer from these formulas the form of the full crosscap state of the model:
$$\left| C\right\rangle=\sum_{j\in\BZ_{k/2}}\root{\scriptstyle 4}\of
{\frac{2}{k+2}}\cot^{1/2}\left(\frac{(2j+1)\pi}{2(k+2)}\right)\left|C,j
\right\rangle,\eqn\eecross$$
where $\left| C,j\right\rangle$ are normalized so as to give the corresponding
character,
$$\left\langle C,j\right| \left.  C,l\right\rangle =\delta _{jl}\chi _j.
\eqn\eechar$$

The open string part of the spectrum is then required to satisfy constraints
\ZMS\  that embody modular properties of the model.  Ref.~[\Ishib] shows that
open strings carrying each of the integer-spin integrable representations at
level $k$ is a good choice.  With the projection operator in the open sector
acting trivially on the primary states, $\Omega\left|j\right\rangle=\left|j
\right\rangle$, the following amplitudes can be computed:
$$\eqalign{Z_{\rm C}(t)&=\sum_{j\in\BZ_{k/2}}\chi_j(\sfrac{it}{2})=\sum_{j\in
\BZ_{k/2},l\in \frac{1}{2}\BZ_k}S^l_j\chi_l(i\tilde{t})\cr
&=\sqrt{\frac{2}{k+2}}\sum_{j\in\BZ_{k/2}}\sin^{-1}\left(\frac{(2j+1)
\pi}{k+2}\right)\chi_j(i\tilde{t}),\cr
Z_{\rm MS}(t)&=\sum_{j\in\BZ_{k/2}}\chi_j(\sfrac{1}{2}+\sfrac{it}{2})=
\sum_{j\in\BZ_{k/2},l\in\frac{1}{2}\BZ_k}M^l_j\chi_l(\sfrac{1}{2}+i
\tilde{t})\cr
&=\frac{1}{\sqrt{k+2}}\sum_{j\in\BZ_{k/2}}\sin^{-1}\left(\frac{(2j+1)
\pi}{2(k+2)}\right) \chi _j(\sfrac{1}{2}+i\tilde{t}),\cr}\eqn\eeampl$$
where the M\"obius strip diagram is transformed to the tree channel by the
modular transformation $M$ acting on the characters as
$$M^i_j=\left\{\eqalign{&\frac{2}{\sqrt{k+2}}\sin\frac{(2j+1)(2i+1)\pi}{2(k+2)}
\qquad{\rm for}\ i+j\in\ \BZ,\cr&0\qquad\qquad\qquad{\rm for}\ i+j\in
\ \BZ+\frac{1}{2}.\cr}\right.\eqn\eecharagain$$
These amplitudes satisfy requirements \ZMS , the boundary state of this model
being
$$\left| B\right\rangle=\sum_{j\in\BZ_{k/2}}\root{\scriptstyle 4}\of
{\frac{2}{k+2}}\sin^{-1/2}\left(\frac{(2j+1)\pi}{(k+2)}\right)\left|B,j
\right\rangle,\eqn\eebound$$
with the states $\left| B,j\right\rangle$ defined analogously as the
$\left|C,j\right\rangle$.

Ishibashi's model was obtained as an orbifold model using the $\ztwo$ action
of \symaction\ on the conventional $\su 2$ WZW model.  Nevertheless, this
worldsheet orbifold model is not equivalent to the model we are searching for,
since its open string spectrum does not correspond to the spectrum of
\opspectrum\ obtained in the $\su 2$ CS gauge theory on orbifolds.

The 2D CFT that does correspond to the $\su 2$ CS gauge theory on orbifolds
as discussed in \S{3} can be identified by inverting the strategy that we used
above when we derived the cylinder and M\"obius strip amplitudes of
Ishibashi's model.  Starting from the ``Chern-Simons inspired'' spectrum
\opspectrum , after some algebra we obtain the one-loop amplitudes
corresponding to this spectrum:
$$\eqalign{Z_{\rm C}(t)&=2\chi_0(t)+2\chi_{k/2}(\sfrac{it}{2})=2\sum_{j\in\{0,
k/2\},l\in\frac 12\BZ_k}S^l_j\chi_l(i\tilde t)\cr
&=4\sqrt{\frac{2}{k+2}}\sum_{j\in\BZ_{k/2}}\sin\frac{(2j+1)\pi}{k+2}\,\chi_j(i
\tilde t), \cr
Z_{\rm MS}(t)&=2\chi_0(\sfrac 12+\sfrac{it}{2})=\sum_{l\in\frac 12\BZ_k}M^l_0
\chi_l(\sfrac 12+i\tilde t)\cr
&=\frac{2}{\sqrt{k+2}}\sum_{j\in\BZ_{k/2}}\sin\left(\frac{(2j+1)\pi}{2(k+2)}
\right)\chi_j(\sfrac 12+i\tilde t).\cr}\eqn\eesp$$
The boundary and crosscap states corresponding to these amplitudes are easily
found to be:
$$\eqalign{\left| B\right\rangle&=\sum _{j\in\BZ_{k/2}}2\root
{\scriptstyle 4}\of{\frac{2}{k+2}}\sin^{1/2}\left(\frac{(2j+1)\pi}{(k+2)}
\right)\left|B,j\right\rangle,\cr\left|C\right\rangle&=\sum_{j\in\BZ_{k/2}}
\root{\scriptstyle 4}\of{\frac{2}{k+2}}\tan^{1/2}\left(\frac{(2j+1)\pi}{2(k+2)}
\right)\left|C,j\right\rangle,\cr}\eqn\eebocr$$
which leads to the Klein bottle amplitude:
$$\eqalign{Z_{\rm KB}&=\sqrt{\frac{2}{k+2}}\sum_{j\in\BZ_{k/2}}\tan
\frac{(2j+1)\pi}{2(k+2)}\,\chi _j(i\tilde{t}) \cr
&=\sqrt{\frac{2}{k+2}}\sum_{j\in\BZ_{k/2},l\in
\frac{1}{2}\BZ_k}\tan \frac{(2j+1)\pi}{2(k+2)}\,
S_j^l\chi _l(2it) \cr
&=\sum_{j\in \frac{1}{2}\BZ_k}(-1)^{2j}\chi _j(2it).\cr}\eqn\eekl
$$
The remarkable simplification of the last formula makes the interpretation of
the Chern-Simons inspired 2D model obvious.  The Klein bottle diagram
corresponds to projecting the $\su 2$ WZW model by a slight modification of
the $\ztwo$ orbifold transformation used in the worldsheet orbifold
interpretation of Ishibashi's model above.  Namely, we now supplement the
orbifold $\ztwo$ action leading to Ishibashi's model, by the action of the
non-trivial central element of $\su 2$:
$$\Omega J^a_n\Omega^{-1}=\bar J^a_n,\qquad\qquad\Omega\left|j\otimes\bar j
\right\rangle=(-1)^{2j}\left|\bar j\otimes j\right\rangle.\eqn\asymaction$$
Hence, the 2D model that we have obtained from the $\su 2$ CS gauge theory can
be interpreted as a worldsheet orbifold model, in which the simplest orbifold
action of \symaction\ is combined with the target action $g\mapsto -g$ on
$\su 2$, a $\ztwo$ mapping known from the context of extended chiral algebras
[\MSZoo].

The one-loop amplitudes of any worldsheet orbifold model should satisfy the
consistency conditions \ZMS .  Possibilities for given amplitudes to satisfy
these constraints depend on the (Chan-Paton) degeneration of open sectors of
the model.  Thus, in the model that corresponds to \asymaction\ and
\opspectrum , each of the two components of the singular locus is equivalent
to $W_0(\ell)+W_{k/2}(\ell)$, and fusion of the two components of the singular
locus produces both components of the open-string Hilbert space, $[\phi_0]$
and $[\phi_{k/2}]$, in duplicate.  This degeneration of the open string
spectrum is the correct degeneration required in order to obey \ZMS .  We have
thus confirmed that in this specific example, the correspondence between CS
gauge theory and 2D worldsheet orbifolds does indeed allow one to identify the
proper Chan-Paton degeneration of the open sector of the theory.
%

\section{Extended Chiral Algebras and 3D Orbifolds}

We have now identified the CFT that corresponds to the $\su 2$ CS gauge theory
on orbifolds.  This CFT is actually a modification of the Ishibashi model of
open strings on the $\su 2$ group manifold.  This makes us wonder whether
Ishibashi's model itself can also be classified with the use of CS gauge
theory on orbifolds.

The only difference between the two worldsheet orbifold models as defined by
\symaction\ and \asymaction\ is a $\ztwo$ twist,
$$\left|j\otimes\bar j\right\rangle\ \rightarrow\ (-1)^{2j}\left|j\otimes
\bar j\right\rangle.\eqn\ptargaction$$
If taken as a $\ztwo$ orbifold action in the $\su 2$ WZW model on closed
oriented surfaces, this $\ztwo$ twist turns the $\su 2$ model into an $\so 3$
WZW model.  According to [\MSZoo], the $\so 3$ WZW model corresponds to the
$\so 3$ Chern-Simons gauge theory on manifolds.  This gives us actually a clue
about the CS gauge theory for Ishibashi's model.  To follow this clue, we
consider the $\su 2$ CS gauge theory, but now we build in the $\ztwo$ twist
difference between \symaction\ and \asymaction\ by fusing each component of
the singular locus with the Wilson line that carries the ``algebra-extending''
representation of spin $\frac k4$.

To confirm that the CFT corresponding to this peculiar CS gauge theory is
indeed the model discovered by Ishibashi, we will quantize the theory on
$\CO_C\times\BR$.  In accord with its definition, our theory now allows for
just those holonomies $h$ around singular points that square to minus one as
elements of $\su 2$.  Consequently, the singular locus can carry just one
representation, of spin $\frac{k}{4}$.  We can infer the spectrum on the
thickened open string from the $\su 2$ fusion rules \fusionrules\ of the
relevant representations carried by the two singular points of $\CO_C$:
$$[\phi_{k/4}]\times[\phi_{k/4}]=\sum_{j\in\BZ_{k/2}}[\phi_j],
\eqn\eesingpts$$
in accord with the structure of Ishibashi's model.

%
\chapterr{More Examples: Chern-Simons Orbifold Zoo}

Moore and Seiberg conjectured an appealing classification [\MSZoo] claiming
that every CFT (or at least target orbifolds and cosets) can be incorporated
into the Chern-Simons approach to CFT.  In the orbifold case, the relevant CS
gauge theories are those with multiply connected gauge groups (for orbifolds
leading to extensions of chiral algebra), and with disconnected gauge groups
of the structure $G\semi\CG$, where the orbifold group $G$ acts on $\CG$ via
automorphisms.  It would be nice to have a similar classification for
worldsheet orbifolds as well.

Before approaching this issue, it will be instructive to extend the class of
examples discussed so far.  Up to now, we have studied Chern-Simons theory
with non-abelian gauge groups that lead to WZW models; in this section, we
will study $c=1$ CFTs and orbifolds thereof.  The structure of these models
will clarify the question of how exotic worldsheet orbifolds can be described
via CS gauge theory.  This question is not only interesting in itself, but
also provides some hints about the incorporation of worldsheet orbifolds into
the classification given by Moore and Seiberg.
%

\section{$\u 1$ Theory and $c=1$ Worldsheet Orbifolds}

CFTs with $c=1$ correspond to $\u 1$, or rather to $\o 2$, CS gauge theory
[\MSZoo].  The $\u 1$ CS gauge theory is an example of the theory with a
multiplyconnected gauge group, and thus corresponds to CFT with an extended
chiral algebra, the chiral algebra of the rational torus [\MSZoo].  Let us
first recall some basic facts about the rational torus.  This model
corresponds to strings in one target dimension $X,\ X\equiv X+2\pi R$.  For
any value of $R$, this model has the $\u 1$ Kac-Moody symmetry.  In our
normalization and notation, the $\widehat{\u 1}$ primaries are
$$
\phi _{m,n}(z,\bar z)=\exp\left\{p_LX_L(z)\right\}\exp\left\{p_RX_R(\bar z)
\right\},\eqn\eepri$$
with left and right momenta
$$\left( p_L,p_R\right)_{\phi _{m,n}} =\left(\frac{m}{2R}+nR,\frac{m}{2R}-nR
\right), \qquad m,n\in\BZ.\eqn\eeleftright$$
For rational values of $2R^2$, say $\frac{p}{q}$, $\phi _{p,q}$ become chiral,
and extend the chiral current algebra of $\u 1$ to the chiral algebra of the
rational torus generated by $\exp\{i\sqrt{2N}X_L(z)\}$ and $\p X(z)$, with
$N=pq$.  The rational-torus CFT contains $2N$ primaries $\phi _r$ of this
chiral algebra:
$$\phi_r=\exp\left\{\frac{ir}{\sqrt{2N}}X(z)\right\},\qquad r=0,1,\ldots 2N.
\eqn\eechi$$
Diagonal modular invariants correspond to $p$ or $q$ equal to one.  (See
[\MSZoo] for details.)

Quantization of the $\u 1$ CS gauge theory on $\CO_C\times\BR$ is quite
analogous to the case of $\su 2$ theory we discussed above.  At level $N$, the
singular locus can carry any representation whose holonomy squares to one.  We
obtain two representations, namely $\phi_0$ and $\phi_N$.  Fusing the
representations carried by the two components of the singular locus according
to fusion rules of the rational torus:
$$[\phi_r]\times[\phi_s]=[\phi_{r+s}],\qquad r,s,r+s\in\BZ_{2N},\eqn\eetorus$$
we get the Hilbert space of $\u 1$ CS gauge theory on $\CO_C$:
$$\CH_{\CO_C}=2\left\{ [\phi _0]\oplus[\phi _N] \right\}.
\eqn\conespect$$
In accord with our discussion in the previous sections, it should be
isomorphic to the spectrum of open states of a $c=1$ worldsheet orbifold.

Worldsheet orbifolds of the (rational) torus were discussed in
[\wso,\WSOdual].  Motivated by the conjectured correspondence with CS gauge
theory, we are mainly interested in $\ztwo$ orbifolds of models with diagonal
modular invariants.  There are essentially two such (classes of) models of
importance to us, one standard and one exotic.  The standard one uses the
worldsheet parity group as the orbifold group, and the exotic one supplements
the parity action by the target reflection,
$$X\mapsto -X.\eqn\targinv$$
(For details, see [\wso,\WSOdual].) These two models are dual to each other,
i.e.\ they are isomorphic up to redefinition $R\rightarrow 1/2R$ of the target
radii.  These two dual pictures of the same system can be used to shed some
light on each other.  In particular, we have a simple geometrical
interpretation of the spectrum of open states of the model [\WSOdual]:  open
strings are (half)-winding states with their ends sitting in either of the two
fixed points of the orbifold involution \targinv .  In particular, this simple
picture elucidates the structure of the (Chan-Paton) degeneration of the open
strings in the model, which is now related to the existence of two fixed
points of \targinv .  As we are now going to see, these results can be
reproduced from quantization of CS gauge theory on orbifolds, where the
Chan-Paton degeneration comes from different ways in which one representation
can be obtained by fusion of the Wilson lines that represent the string
boundaries in CS gauge theory.

Let us first consider the ``standard'' $c=1$ worldsheet orbifold at
$R^2=1/2N$, which corresponds to a diagonal modular invariant of the orbifold
model.  The open spectrum of the model contains states with momenta
$$p_{\rm open}=\frac{m}{2R}\equiv \frac{m\sqrt{2N}}{2}.\eqn\eetwmom$$
The spectrum can be decomposed into two irreducible representations of the
symmetry algebra of the orbifold model, depending on whether $m$ is even or
odd.  These two representations are exactly the two representations of
\conespect .  Consequently, the $\u 1$ CS gauge theory on orbifolds as
discussed in \conespect\ corresponds to the ``standard'' $c=1$ worldsheet
orbifold at radius $R=\frac{1}{\sqrt{2N}}$.  In particular, the factor of two
in front of the r.h.s.\ of \conespect\ gives the Chan-Paton degeneracy and
counts the fixed points of \targinv\ in the dual picture of the model.

At $R^2=N/2$, the spectrum of open states of the standard worldsheet orbifold
model carries momenta
$$p_{\rm open}=\frac{m}{2R}\equiv \frac{m}{\sqrt{2N}},\eqn\tttwo$$
and can be decomposed into $2N$ representations of the symmetry algebra of the
model.  To get a CS gauge theory description of this region of large target
radii, we will proceed in analogy with the analysis of the $\su 2$ worldsheet
orbifolds in the previous section.  We know from the duality mentioned above
that we are looking in fact for a Chern-Simons interpetation of the exotic
worldsheet orbifold at $R^2=1/2N$.  Thus, we will construct a new $\u 1$ CS
gauge theory on orbifolds as follows.  We supplement the $\ztwo$ action on
manifolds by a $\ztwo$ twist, which is a Chern-Simons analogy of the target
transformation \targinv :  Over the singular locus, we extend the orbifold
group to $\o 2=\ztwo\semi\u 1$ and allow only those holonomies that do not
belong to the $\u 1$ subgroup in $\o 2$; all other holonomies take values in
$\u 1$.  This prescription defines a gauge theory, in which the singular
points of $\CO_C$ are now labeled by twisted primaries of the chiral algebra
of the $\ztwo$ (target) orbifold [\DVVV].  Using the relevant fusion
rules:\foot{Our notation here is that of [\DVVV].}
$$\eqalign{[\sigma_i]\times[\sigma_i]&=[1]+[\phi^i_N]+\sum_{r\ {\rm even}}[
\phi_r],\cr
[\tau_i]\times[\tau_i]&=[1]+[\phi^i_N]+\sum_{r\ {\rm even}}[\phi_r],\cr
[\sigma_i]\times[\tau_i]&=[j]+[\phi^{i+1}_N]+\sum_{r\ {\rm even}}[\phi_r],\cr}
\qquad
\eqalign{[\sigma_1]\times[\sigma_2]&=\sum_{r\ {\rm odd}}[\phi_r],\cr
[\tau_1]\times[\tau_2]&=\sum_{r\ {\rm odd}}[\phi_r],\cr
[\sigma_i]\times[\tau_{i+1}]&=\sum_{r\ {\rm odd}}[\phi_r],\cr}\eqn\eefus$$
and composing the result into representations of the rational-torus chiral
algebra, one gets exactly the spectrum \tttwo\ of the $2N$ representations of
the standard worldsheet orbifold at $R^2=N/2$, with a Chan-Paton degeneration
of the open sector.  Thus, the exotic worldsheet orbifold at small radius, or
alternatively the standard orbifold at large radius, corresponds to the
twisted $\u 1$ CS gauge theory we have just constructed.

Two interesting consistency checks can be made immediately.  First, notice
that using $\o 2$ as the gauge group on 3D orbifolds, one can easily recover
the model that corresponds to open strings on target orbifold $S^1/\ztwo$,
which mixes in the obvious way the two $c=1$ models just discussed, producing
simultaneously the $\ztwo$-twisted closed states, as necessary.  Secondly,
notice that the compact boson at the self-dual radius, $R=1/2$, can be
reconstructed from each of the two CS gauge theories presented above.  It is
reassuring that both descriptions give the same result.  This closes our study
of $c=1$ worldsheet orbifolds via CS gauge theory.
%

\section{Discrete Gauge Groups}

Many crucial points of the relation between CFTs on surfaces with boundaries
and/or crosscaps (i.e.\ worldsheet orbifold models) on one hand and the 3D
CS gauge theory on the other can be efficiently isolated by studying
holomorphic orbifolds [\DVVV].  At the level of CS gauge theory, holomorphic
orbifolds are described by discrete gauge groups [\DijkW,\MSZoo].

Let us consider the CS gauge theory with an arbitrary finite gauge group $G$,
on $\ztwo$ orbifolds.  I will limit the discussion to the classical theory;
quantization can be treated similarly as in [\DijkW], after a choice of an
element of $H^4_\ztwo(B(\ztwo,G),\BZ)$ is made, to represent the choice of a
Lagrangian (see Appendix B).

The phase space for canonical quantization on an orbifold $\CO$ is given by
the space of flat principal $G$-bundles over $\CO$.  On the thickened open
string $\CO_C$, flat principal $G$-bundles are classified by representations
of the fundamental group $\pi_1(\CO_C)$ in the gauge group:
$$
\homotopy \rightarrow G.\eqn\eehomotopy$$
This recovers the picture of standard worldsheet orbifolds in two dimensions
as discussed is \S{2.1}:  Flat $G$-bundles over $\CO_C$ are in one-to-one
correspondence with the monodromies \standard\ of the fields on the open
string, i.e.\ they are in one-to-one correspondence with the open twisted
states of a standard worldsheet orbifold (cf.\ \S{2}).

Exotic worldsheet orbifolds can also be obtained in a simple way.  To
construct the CS gauge theory corresponding to a (holomorphic) exotic orbifold
with orbifold group $G\subset \widetilde G\times\wsgroup$, we have to sum over
a restricted class of $\widetilde G$-bundles over $\ztwo$ orbifolds.  This
restriction corresponds to the commutativity restriction discussed in \S{2.1}
(cf.\ \tobetriangle).  Upon denoting by $G_0$ the set of all the elements from
$G$ that act trivially on the worldsheet and thus represent a target orbifold
group, $G$ can be written as $\ztwo\semi G_0$.  Given now an orbifold $\CO$,
its fundamental group $\pi_1(\CO)$ has the structure of a $\ztwo$ extension of
$\pi_1(\bar\CO)$, the fundamental group of its double cover $\bar\CO$ (which
is, by assumption, a manifold).  The allowed holonomies are now required to
respect these $\ztwo$ extensions on $G$ and $\pi_1(\CO)$, i.e., they should
make the following diagram commutative:
$$\matrix{1&\rightarrow&\pi_1(\bar\CO)&\rightarrow&\pi_1(\CO)&\rightarrow&\ztwo
&\rightarrow&1\cr
&&\downarrow&&\downarrow&&\downarrow{\scriptstyle\rm id}&&\cr
1&\rightarrow&G_0&\rightarrow&G&\rightarrow&\ztwo&\rightarrow&1\cr}
\eqn\exactseq$$
This commutative diagram thus defines a variant of CS gauge theory, in which
the gauge group is intertwined non-trivially with the action of the orbifold
group $\ztwo$.  Hence, the exotic holomorphic worldsheet orbifolds can be
given a three dimensional Chern-Simons description, with the exact form of the
allowed holonomies is encoded in the requirement of commutativity of
\exactseq .  I will speculate on the nature of this exotic version of gauge
theory in the following subsection.
%

\section{Gauging a Mapping Class Group}

In the previous subsections we have seen examples of worldsheet orbifold CFTs
whose corresponding CS gauge theories apparently intertwine in a non-trivial
way the gauge group with the action of the orbifold group $\ztwo$.  For
example, the CS gauge theory that was shown to correspond to \tttwo\ is
neither an $\o 2$ gauge theory, nor a $\u 1$ theory on $\ztwo$ orbifolds,
since the holonomies around singular points are treated differently from the
holonomies around the non-contractible circles of the conventional manifold
origin.  I will now argue that the proper way how to interpret such a theory
is to think of $\ztwo$ as a part of the gauge group, in a very specific sense
that amounts to gauging the $\ztwo$ as a part of the mapping class group of
the underlying manifold.

Before discussing the construction, however, I will present yet another
heuristic argument that $\wsgroup$ should be treated as a part of the gauge
group.  It is shown in Appendix B that consistent Lagrangians for CS gauge
theory on $\ztwo$ orbifolds are classified by elements of the fourth $\ztwo$
equivariant cohomology ring $H^4_\ztwo(\classif,\BZ)$, where $\CG$ is the
original Chern-Simons gauge group, and $\classif$ is tom~Dieck's classifying
space of principal $\CG$-bundles over $\ztwo$-manifolds.  Recalling that
$$H^{\ast}_\ztwo(\classif,\BZ)=H^{\ast}(B(\ztwo\times\CG),\BZ),
\eqn\eeclasshom$$
(see Appendix B), it is easy to see that the right hand side is exactly the
object that classifies consistent Lagrangians for the gauge group
$\ztwo\times\CG$ on manifolds.  We might interpret this fact as a signal that
the $\ztwo$ group acting on manifolds has become a part of the gauge group.

We are free to define a gauge theory that corresponds to the ideas presented
above, as follows.  In the standard construction of a gauge theory whose gauge
fields correspond to connections $A$ on a principal $\CG$-bundle, the
functional integral that defines the theory on a given manifold $M$ contains
summation over all principal $\CG$-bundles on $M$:
$$Z(M)=\sum_{\CG{\rm -bundles}}\int_\CA DA\;e^{i\CS(A)},\eqn\instantonsum$$
where $\CA$ denotes the set of gauge equivalence classes of the connection,
$\CS(A)$ is a gauge invariant Lagrangian, and a gauge fixing procedure which
gives sense to the formal functional integral is implicitly assumed.  In
paricular, for finite gauge groups [\DijkW,\KWilcz] this summation
distinguishes the gauge theory from the theory with the discrete symmetry
being just global.

We can now define the theory with a mapping class group gauged, as a simple
extension of the construction just discussed.  Let us consider a $\ztwo$
extension of $\CG$, $\ztwo\semi\CG$.  With this as a gauge group, the sum in
\instantonsum\ would run over all $\ztwo\semi\CG$ principal bundles.
Recalling that principal $\ztwo\semi\CG$ bundles are spaces with free
$\ztwo\semi\CG$ actions, we will now modify the sum so as it will now run
over all $\ztwo\semi\CG$-spaces that are $\CG$-free, but not necessarily
$\ztwo\semi\CG$-free.  In other words, these spaces can be thought of as
(total spaces of) principal $\CG$-bundles with a $G$-action on them.  Thus,
instead of summing over principal bundles classified by the classifying space
$B(G\times\CG)$, which would correspond to the conventional $\ztwo\semi\CG$
gauge theory, we are summing in the ``exotic'' version of gauge theory over
the objects classified by tom~Dieck's classifying space $B(G,\CG)$.  In this
case, $\ztwo$ acts on 3D manifolds as an element of their mapping class group,
which explains the title of this subsection.  The CS gauge theory on $\ztwo$
orbifolds as discussed in the previous sections confirms that this approach
really makes sense, since the theory represents a concrete example of the
formal definition of the ``exotic'' gauge theory.  In particular, the CS gauge
theories discussed in \S{5.1} and \S{5.2} are examples of gauge theories of
this type.  Hence, this extension of the standard definition of CS gauge
theory allows us to add the CFTs of worldsheet orbifolds to Moore and
Seiberg's list of 2D conformal field theories classifiable by their
corresponding 3D CS gauge theories.

%
\chapterr{Concluding Remarks}

In this paper I have studied CS gauge theory on three-dimensional $\ztwo$
orbifolds.  This theory is interesting not only because it satisfies the
axioms of equivariant topological field theory [\etsm], but especially because
it is intimately related to 2D CFT on surfaces with boundaries and/or
crosscaps (the so-called ``worldsheet orbifold'' CFTs).  This relation gives
us new insight into several aspects of open string theory; it may also have
implications for the boundary conformal scattering in two dimensions, a
problem that itself has many interesting ramifications [\affleck].

We have seen that the 2D and 3D aspects of CS gauge theory on orbifolds
illuminate each other in an interesting way:

For one, the three dimensional description of 2D CFT reveals the geometrical
origin of the Chan-Paton mechanism (responsible for the existence of spacetime
Yang-Mills gauge symmetry in open string theory).  {}From this point of view,
the rationale for the existence of the Chan-Paton symmetry is in
three-dimensional algebraic topology, namely in the existence of twisted
principal $\CG$-bundles of the corresponding Chern-Simons gauge group $\CG$
(not to be confused with the resulting Chan-Paton gauge group) on three
dimensional $\ztwo$ orbifolds.  Moreover, our interpretation of the open
string boundary as a link of Wilson lines in CS gauge theory leads to a
surprisingly simple prescription for the identification of open string spectra
in 2D CFTs, in terms of fusing specific Wilson lines in CS gauge theory.

On the other hand, the analysis of several specific 2D CFTs (related to the
so-called ``exotic worldsheet orbifolds'') in fact suggests the existence of a
more unified treatment for the three dimensional gauge theory, in which the
original gauge group and the orbifold $\ztwo$ group become two parts of a
larger gauge group.  Conceptually, this step can be considered an extension of
the standard definition of gauge theories.  This extended class of 3D CS gauge
theories then allows us to extend the classification results of [\MSZoo] to
open string theory, proving in particular that at least those open string
models that can be interpreted as worldsheet orbifold models in the sense of
\S{2} do fit into the general classification by Moore and Seiberg that uses
three-dimensional CS gauge theory to classify 2D CFTs.

Note also that in this paper, we have constructed a quantum field theory on
spaces with singularities.  This might be particularly interesting when
combined with studies of $2+1$-dimensional quantum gravity which can be
formulated as CS gauge theory with a non-compact gauge group
[\Wgrav,\Carlip,\Wtrans].  In fact, $2+1$-dimensional quantum gravity on
orbifolds (not necessarily $\ztwo$ ones) might give us an exactly soluble
quantum theory with (mild) spacetime singularities completely under control.
In general, the effective equivalence between orbifold singularities and
Wilson lines as seen in this paper may be considered a toy example of the
idea that black holes (represented by the singular locus in this toy example)
have just as much of hair as any particle [\KWilcz], simply because the
singularities are equivalent to a sum over Wilson lines that represent
physical particles.

This paper is just a small step towards the complete CS gauge theory on
orbifolds and its full correspondence with 2D CFT on surfaces with boundaries
and/or crosscaps.  The reader undoubtedly noticed that I have frequently
chosen the way of smallest resistance instead of considering the most general
situation possible.  Indeed, the focus of this paper has been on the main line
of arguments that leads as effectively as possible from CS gauge theory on
orbifolds to CFTs of worldsheet orbifolds, and allows us to discuss specific
examples.  Many interesting aspects of the story had to be left out for future
investigation.

%
\APPENDIX{A}{ A: Geometry of Three-Dimensional Orbifolds}
\message{Appendix A: Geometry of Three-Dimensional Orbifolds}

The Chern-Simons gauge theory is a theory of (flat) connections on
three-dimensional ``spacetimes.''  To be able to study the theory on
``spacetimes'' that are orbifolds, we need some basic elements of orbifold
geometry.

An $n$-dimensional orbifold is defined as a space modelled locally by factors
of domains in $\BR^n$ by discrete groups.  More precisely, we will define an
orbifold $\CO$ as an underlying Hausdorff topological space $X_\CO$ with a
maximal atlas of coverings by open sets $\{ U_i\}$.  If $\CO$ were a manifold,
the $U_i$s would be open subsets in $\BR^n$.  In the case of orbifolds, we
associate with each $U_i$ a discrete group $G_i$, such that $U_i$ is a factor
of a domain $\bar U_i\subset\BR^n$ by $G_i$,
$$U_i=\bar U_i/G_i, \eqn\eeaone$$
(To avoid some counter-intuitive cases, we require that $G_i$ act on
$\bar U_i$ effectively.)  Maps between charts are required to respect the
group action.

For each point $x$ in an orbifold $X_\CO$, the smallest group $G_i$ associated
to a domain containing $x$ is called the ``isotropy group'' of $x$.  The
subset in $X_\CO$ of points whose isotropy group is non-trivial is called the
locus of singular points, or the ``singular locus'' of $\CO$.

To be able to define fibered bundles over orbifolds, a structure that we need
in gauge theory, we must first define the notion of morphisms between
orbifolds.  A ``morphism'' from orbifold $\CO$ to another orbifold $\CO'$ is
defined as a mapping $f$ between the underlying spaces, $f: X_\CO\rightarrow
X_{\CO'}$, that respects the orbifold structure of $\CO$ and $\CO'$, i.e.\
it respects the group action in each coordinate chart.  As a consequence,
if $x$ is an arbitrary point in $\CO$ with isotropy group $G_x$ and $y=f(x)$,
then necessarily the isotropy group $G_y$ of $y$ contains $G_x$ as a
subgroup.

This definition of morphisms between orbifolds gives us a category of
orbifolds, in which such notions as covering maps, fibered bundles, homotopies
between maps, homotopy groups etc.\ can be straightforwardly defined.  For
example, the mapping that retracts the thickened open string $\CO_C$ to the
open string $\CO_S$ itself (see figure~(2)) represents a homotopy from $\CO_C$
to $\CO_S$ (in fact, this map is a deformation retraction; for the definition
of the latter in the case of manifolds, see e.g.\ [\Spanier]).  This fact
explains the observation made in the paper that the orbifold fundamental
groups of $\CO_C$ and $\CO_S$ are isomorphic.

The category of orbifolds is very similar to the category of $G$-spaces with
$G$-equivariant maps as morphisms (here $G$ is an {\it a priori} fixed finite
group).  For the purposes of this paper, these two categories can be
considered in many respects equivalent.

Let us now proceed from the topology of orbifolds to their geometry.  We
define a principal $\CG$-bundle for any (Lie) group $\CG$ over an orbifold
$\CO$ as follows.  Let $P$ be an orbifold fibered over an orbifold $\CO$
(i.e.\ the projection $\pi:\CP\rightarrow\CO$ is an orbifold morphism), such
that for each chart $U_i$ on $\CO$ one is given a representation of $G_i$ in
$\CG$, and for $U_i$ from a suficiently refined covering of $\CO$, we have
$$\pi^{-1}(U_i)=(\bar U_i\times\CG)/G_i,\eqn\eeatwo$$
where the action of $G_i$ on $U_i$ is that of \eeaone , and the action on
$\CG$ is given by the representation of $G_i$ in $\CG$.  Then $\CP$ is what
we can call the total space of a principal $\CG$-bundle over $\CO$.

To be more specific, let us illustrate the definition of the principal bundle
by classifying principal $\su 2$ bundles over our favorite orbifold $\CO_C$.
To construct an $\su 2$ principal bundle over $\CO_C$, we have to specify a
representation of the orbifold group $\ztwo$ in $\su 2$, over each singular
point in $\CO_C$.  There are two such representations possible, over each
singular point.  One of them is trivial and maps $\ztwo$ to the identity in
$\su 2$, and the other one maps $\ztwo$ to the center of $\su 2$.  These two
representations represent two possible twists of a principal $\su 2$ bundle
over a singular point of $\CO_C$.  One of them gives a topologically trivial
bundle over a vicinity of the singular point, while the other one is
``twisted,'' and effectively reduces the structure group over the singular
point from $\su 2$ to $\so 3$.  Note the amusing fact that in the twisted
case, the total space of the principal bundle over a vicinity of the singular
point is a manifold, and the singularity of the bundle is due to a singular
projection to the base orbifold $\CO_C$.

%
\APPENDIX{B}{ B: Lagrangians of CS Gauge Theory on Orbifolds}
\message{Appendix B: Lagrangians of CS Gauge Theory on Orbifolds}

In this Appendix I present some technicalities of the definition of
Chern-Simons Lagrangians for general (compact) gauge group, not necessarily
connected or simply connected, on three-dimensional $\ztwo$ orbifolds.  The
analysis follows closely the non-equivariant case discussed in [\DijkW].

Regardless of what the gauge group is, the requirements of factorization in
the theory on orbifolds make the Lagrangian $S(A)$ on $\CO$ zero if $\CO$ is
the boundary of a four-dimensional $\ztwo$ orbifold $\CB$ assuming $A$ can
be extended as a flat connection over $\CB$.  First we will check whether
there exists an obstruction for a three dimensional $\ztwo$ orbifold to be the
boundary of a four-dimensional $\ztwo$ orbifold.  If such an obstruction
existed, it would be an element of the third $\ztwo$ equivariant cobordism
group $I_3(\ztwo)$ (see [\CFloyd]).  In general, $I_{\ast}(\ztwo)$ is defined
as the group of equivalence classes of $\ztwo$ manifolds, two of them being
equivalent if they bound a $\ztwo$ manifold.  Using a split exact sequence
[\CFloyd], the cobordism group of our interest can be easily calculated,
leading to $I_3(\ztwo)=\ztwo$.  Hence, there is a $\ztwo$ obstruction for some
three dimensional $\ztwo$ orbifolds to represent the boundary of a
four-dimensional $\ztwo$ orbifold, which indicates that the definition of
Lagrangian must be treated carefully.  To this goal, we will modify to the
orbifold case the results of [\DijkW], where the Lagrangian for CS gauge
theory on manifolds has been defined using group cohomology.

Let us consider CS gauge theory on $\ztwo$ orbifolds with a compact gauge
group $\CG$, not necessarily connected or simply connected.  In the case of
the theory on manifolds, the principal bundles over a given manifold $M$ are
classified by homotopy classes of mapping of $M$ to the ``classifying space''
$B\CG$, and the consistent Lagrangians are classified by the fourth cohomology
group $H^4(B\CG,\BZ)$ [\DijkW].

For principal bundles over orbifolds, relevant classifying spaces were defined
and studied by tom~Dieck in [\Dieck] (see also [\Stong]).  His classifying
space $\classif$ has the property that for any principal $\CG$-bundle $E$ over
a manifold $B$ with a $\ztwo$ action on $E$ and $B$ commuting with the
$\CG$-action on $E$, there exists a $\ztwo$ equivariant mapping (unique up to
$\ztwo$ equivariant homotopy) of $B$ to $\classif$, which induces $E$ on
$B$ from a universal bundle over $\classif$.%
\foot{Ref.\ [\Dieck] discusses a generalization of this construction to the
case of general semi-direct products $G\semi\!{}_{\alpha\;}\CG$ as well, i.e.\
to principal $\CG$-bundles with a $G$-action commuting with the $\CG$-action
on the total space up to a representation $\alpha$ of $G$ in the group of
$\CG$-automorphisms.  The corresponding classifying spaces, denoted as
$B(G,\alpha,\CG)$, are relevant to the ideas of \S{5.3} of the paper.}
I now claim that the consistent Lagrangians of the CS gauge theory on
orbifolds are classified by elements of the fourth equivariant cohomology
[\Bredon] of tom~Dieck's classifying space, $H^4_\ztwo(\classif,\BZ)$.

To show this, let us first compute the relevant cohomology group.  As shown in
[\Dieck], the classifying space $B(G,\CG)$ is homotopic to the classifying
space $B\CG$, once the action of $G$ on $B(G,\CG)$ is ignored.  Thus, we can
find a representant of $B\CG$ such that $G$ acts on it, and
$$H^{4}_\ztwo(\classif,\BZ)=H^{4}_\ztwo(B\CG,\BZ).\eqn\eebone$$
With the use of the definition of equivariant cohomologies, we easily obtain
$$\eqalign{H^{\ast}_\ztwo(\classif,\BZ)&=H^{\ast}((B\CG\times E\ztwo)/\ztwo,
\BZ)= H^{\ast}(((E\CG/\CG)\times E\ztwo)/\ztwo,\BZ)\cr
&=H^{\ast}((E\CG\times E\ztwo)/(\ztwo\times\CG),\BZ)=H^{\ast}(B(\ztwo\times
\CG),\BZ).\cr}\eqn\eebtwo$$
Using now the K\"{u}nneth formula [\BottTu] for integral cohomologies, we get
the following result for the fourth cohomology group of our interest:
$$H^{4}_\ztwo(B\CG,\BZ) = H^4(B\CG,\BZ)\oplus H^2(B\CG,\ztwo)\oplus
\ztwo.\eqn\ccohomology$$
Let us now consider a principal bundle $E$ over an orbifold $\CO$.  Its double
covering $\bar E$ over $\CO$ is an example of the objects classified by
$\classif$.  Let $B$ is a four-manifold with the boundary $\bar\CO$.  Given a
classifying map $\bar\CO\rightarrow\classif$ of $E$, there is an obstruction
to extending it to an equivariant mapping $B\rightarrow\classif$, given by the
element $\gamma_\ast[\bar\CO]$ of the third $\ztwo$ equivariant homology group
$H_3^\ztwo(\classif,\BZ)$.  The torsion part of this group is isomorphic by
the universal coefficients theorem to the torsion of $H^{4}_\ztwo(\classif,
\BZ)$ (see \ccohomology ).  Supposing for simplicity that the torsion of
$H^4(B\CG,\BZ)$ is $\BZ_n$, then $2n\cdot\gamma_\ast[\bar\CO]$ is
homologically trivial in $\classif$.  We can thus define the Lagrangian on
$\CO$ by
$$2n\cdot S(A)=\frac{k}{8\pi}\int _P{\rm Tr}(F\wedge F),\eqn\eebfour$$
where $\p P$ consists of $2n$ copies of $\bar\CO$.  A resolution of the
$2n$-fold ambiguity of this definition is then given, recalling that the form
$\Omega \equiv \frac{k}{8\pi ^2}{\rm Tr}(F\wedge F)$ is in the image of the
natural map $H^4_\ztwo(\classif,\BZ)\rightarrow H^4(B\CG,\BR)$, by any element
of the fourth equivariant cohomology of $\classif$ as claimed above.

To summarize, the consistent Lagrangians for CS gauge theory with gauge group
$\CG$ on $\ztwo$ orbifolds are classified by the fourth $\ztwo$ equivariant
integer cohomology group of the classifying space $\classif$.  For the
purposes of the present paper (cf.\ \S{5.3}), the most relevant point is
that this cohomology group is isomorphic to the fourth integer cohomology
group of the classifying space of $\ztwo\times\CG$.

\message{* References}
\refout
\end
\bye